\documentclass[useAMS,usenatbib]{mn2e}
\usepackage{graphicx}
\usepackage{epstopdf}
\usepackage[figuresright]{rotating}
\usepackage{subfigure,color}
\usepackage{longtable}
\usepackage{tabularx,multicol,lipsum}
\usepackage{xspace}

\newcommand{\ion}[2]{{\textrm{#1}}{\textrm{\sc #2}}}

\definecolor{pink}{rgb}{.9,.2,.5}  
\definecolor{purple}{rgb}{.5,.6,.7}

\title[Semi-empirical calibrations]
{Semi-empirical metallicity calibrations based on 
ultraviolet  emission lines of type-2 AGNs}

\author[Dors et al.]
            {O.~L.\ Dors$^{1}$\thanks{E-mail:olidors@univap.br}, A.~F.~Monteiro$^{1}$, M.~V.~Cardaci$^{2,3}$, 
	    G.~F.~H\"{a}gele$^{2,3}$, A.~C. Krabbe$^{1}$
               \\
$^1$ Universidade do Vale do Para\'iba, Av. Shishima Hifumi, 2911, Cep
12244-000, S\~ao Jos\'e dos Campos, SP, Brazil\\ 
$^{2}$ Instituto de Astrof\'{i}sica de La Plata (CONICET-UNLP), Argentina \\
$^{3}$ Facultad de Ciencias Astron\'{o}micas y Geof\'{i}sicas, Universidad Nacional de La Plata, Paseo del Bosque s/n, 1900 La Plata, Argentina
} 
\begin{document}

\date{Accepted 2015 Month  00. Received 2015 Month 00; in original form 2014 December 17}

\pagerange{\pageref{firstpage}--\pageref{lastpage}} \pubyear{2011}

\maketitle

\label{firstpage}

\begin{abstract}
We derived two semi-empirical calibrations between the metallicity of the Narrow Line Region (NLR)
of type-2 Active Galactic Nuclei
and the rest-frame of the \ion{N}{v}$\lambda$1240/\ion{He}{ii}$\lambda1640$, 
C43=log[(\ion{C}{iv}$\lambda1549$+\ion{C}{iii}]$\lambda1909$)/\ion{He}{ii}$\lambda1640$]
and \ion{C}{iii}]$\lambda1909$/\ion{C}{iv}$\lambda1549$  emission-line intensity ratios.  A 
metallicity-independent calibration between the ionization parameter and 
the \ion{C}{iii}]$\lambda1909$/\ion{C}{iv}$\lambda1549$ emission-lines ratio was also derived.
These calibrations were obtained comparing  ratios of measured  UV emission-line intensities, compiled from the literature, for 
a sample of 77 objects (redshift   $0 \: < \: z \: < \: 3.8$) with those predicted by a grid of photoionization models built with 
the {\sc Cloudy} code. 
Using the derived calibrations, it was possible to show that the metallicity estimations for NLRs are lower
 by a factor of about 2-3 than those for Broad Line Regions (BLRs). Besides we confirmed the recent  result 
 of the existence of a relation between the  stellar mass of the host galaxy and its NLR metallicity.
  We also derived a $M-Z$ relation for the objects in our sample at $1.6 \: < \: z \: < \: 3.8$. This relation 
seems to follow the same trend as the ones estimated for Star Forming galaxies of similar high redshifts but for higher masses.
\end{abstract}



\section{Introduction}
\label{intro}

Active Galaxy Nuclei (AGNs) present prominent metal emission-lines in their spectra, easily measured 
in a wide range of wavelengths even when the objects are located at very high redshift ($z \: > 5$). The intensity 
of these lines can be used to estimate the gas phase metallicity, becaming  AGNs essential in studies 
of chemical evolution of galaxies as well as of the Universe.

In general, it is assumed that  fiducial metallicity ($Z$) estimations of  emitter line objects 
(e.g.\ AGNs, \ion{H}{ii} regions, and Planetary Nebulae) are those obtained through the $T_{\rm e}$-method
 \citep{osterbrock06} which, basically, consists of using  collisionally-excited forbidden emission-lines to estimate
  the electron temperature and the abundance of a given element in relation to the hydrogen abundance, usually O/H. 
  However, \citet{dors15} showed that oxygen abundances, estimated from  the $T_{\rm e}$-method 
   and  from narrow emission lines of a sample of  Seyfert 2 AGNs,
are underestimated  by up to $\sim 2$ dex (with averaged value of  $\sim 0.8$ dex) in relation to the expected values
  from the extrapolation of  radial abundance gradients. 
These authors also showed that  estimations of metallicity based on theoretical calibrations 
of strong optical emission  lines seem to be reliable for AGNs.

Most of the metallicity calibrations for  AGNs (\citealt{thaisa98, castro17}) for  \ion{H}{ii} regions 
 (e.g. \citealt{pagel79, edmunds84, pilyugin00, pilyugin01, kewley02, dors05, stasinska06, maiolino08, berg11, 
 perezmontero14, brown16, pilyugin16})  and even for diffuse ionised gas \citep{kumari19}
 are  based on strong optical emission lines 
(e.g.\ [\ion{O}{ii}]$\lambda$3727, H$\beta$, [\ion{O}{iii}]$\lambda$5007, [\ion{N}{ii}]$\lambda$6584, H$\alpha$).
These calibrations, together with the large amount of optical  spectroscopic data obtained by several surveys, such as 
the Calar Alto Legacy Integral Field Area (CALIFA) survey \citep{sanchez12} and the Sloan Digital Sky Survey \citep[SDSS;][]{york00},  
have  revolutionized the extragalactic astronomy. However, optical emission-lines shift out of the
$K$-band atmospheric window for objects at high redshifts ($z \: > 3.5$), requiring space-based spectroscopic data to access
to the most distant objects \citep{matsuoka18, maiolino18}. 
To circumvent this limitation in using optical emission lines, ultraviolet (UV; $\rm 1000\AA \: < \: \lambda \: < 2000\AA$) 
lines can be used to estimate metallicities in a wide redshift range \citep[e.g.][]{nagao06a, matsuoka09, dors14, perezmontero17,
 mignoli19}.
Moreover, UV narrow lines are not significantly affected by gas shocks \citep{matsuoka09}  which
even with low velocities ($v \la 400$ km/s) are 
present in the narrow line regions (NLRs) of AGNs \citep{dors15, contini17}. 

One of the first abundance estimations in AGNs using UV emission-lines was carried out by 
\citet{shields76}, who derived the N/C abundances ratio for the Quasar PKS\,1756+237 located at a redshift of  \textit{z}=1.72.
Later, several authors \citep[e.g.][]{davidson77, osmer80, uomoto84, gaskell81, hamann92, ferland96, dietrich00, 
hamann02, nagao06a, shin13, dors14, feltre16, yang17, dors18} have estimated metallicities and/or elemental abundances (usually nitrogen abundances)
 in Quasars or Seyferts. These estimations have been mainly based on detailed photoionization models \citep[e.g.][]{ferland96, dors18} or 
 made through diagrams containing observational and model predicted emission line intensities \citep[e.g.][]{nagao06a, matsuoka09, matsuoka18},
  being these methodologies difficult to apply for a large sample of data. In this sense, general calibrations
between metallicity and  strong emission-lines or  Bayesian approach \citep{mignoli19}
are preferable and easily applicable.

\citet{dors14}, using photoionization model results, proposed the first calibration between the metallicity of the gas in the NLRs
 of AGNs and the intensity of UV emission-lines through the use of the 
 C43=log[(\ion{C}{iv}$\lambda1549$+\ion{C}{iii}]$\lambda1909$)/\ion{He}{ii}$\lambda1640$] emission-lines ratio.
This calibration is strongly dependent on the ionization degree of the emitting gas which, once the ionizing spectrum is 
 determined,  can be estimated from the \ion{C}{iii}]$\lambda1909$/\ion{C}{iv}$\lambda1549$ emission-lines ratio \citep{davidson72}.  
However, as was pointed out by \citet{dors14}, this \ion{C}{iii}]/\ion{C}{iv} ratio is somewhat depend on the metallicity, 
mainly for low  ionization values. 

Recently, \citet{castro17} presented a new methodology to calibrate the optical 
$N2O2 \rm = \log ([\ion{N}{ii}]\lambda 6584/ [\ion{O}{ii}] \lambda 3727)$ emission-line ratio 
with the metallicity of NLRs of AGNs, producing a semi-empirical calibration. This methodology 
consists of calculating the metallicity ($Z$)  and the corresponding line ratio for a sample of type 2 AGNs through 
diagnostic diagrams containing both observational data and photoionization model results. This method is based on 
the idea proposed by \citet{pilyugin00, pilyugin01}, 
 which consists in  obtaining calibrations  using oxygen abundances (or metallicities) derived from direct 
electron temperature estimations.  The $T_{\mathrm{e}}$-method seems does not work for AGNs \citep{dors15},
probably due to the necessity of applying an Ionization Corretion Factor to the oxygen abundance determinations through this method 
(not explored up to now in this kind of objects), and the probable presence of electron temperature fluctuations and/or shocks 
(neither considered in our photoionization models).  However, \citet{dors15} showed that O/H derived from the 
strong emission-line calibrations 
are in agreement with those (independent) estimations from abundance gradients extrapolations. Thus, the metallicity 
and the physical conditions of the gas
  would seem to be better determined
 applying  (semi) empirical calibrations developed
 through photoionization models,  despite  these are  subject to several uncertainties (e.g.\ \citealt{viegas02, kenniccutt03}). 
 
 The methodology proposed by \citet{castro17} has the advantage of obtaining calibrations considering the closest modeled 
  physical conditions to the real ones  for each objects parametrized through diagnostic diagrams and, therefore,  reducing 
  the uncertainties in  metallicity estimations. 
In fact, for example, for \ion{H}{ii} regions it is well known that, in general,  oxygen abundances estimated 
from purely theoretical calibrations are overestimated in comparison to those derived using the $T_{\mathrm{e}}$-method 
\citep[e.g.][]{kenniccutt03, dors05}. However, \citet{dors11} showed that this discrepancy can be alleviated if 
it is required that the  photoionization models simultaneously reproduce observational line-ratios sensitive to the
 metallicity and to the ionization degree of \ion{H}{ii} regions \citep[see also][]{dors05, morisset16}.

Up to now,  the unique metallicity calibration for AGNs based on narrow ultraviolet line ratios seems to be the theoretical
 one proposed by \citet{dors14}, which is based on the C43 index. It is possible to obtain calibrations involving other 
 strong UV line ratios, such as the \ion{N}{v}$\lambda$1240/\ion{He}{ii}$\lambda1640$ ratio, suggested by \citet{ferland96}
  as a metallicity indicator.

Keeping the above in mind, we combined observational intensities of UV narrow emission-lines of type-2 AGNs compiled from the 
literature with  photoionization model results  to obtain  semi-empirical calibrations between the metallicity of the gas phase
 and the  \ion{N}{v}/\ion{He}{ii} emission-line intensity ratio. We also re-calibrated the C43 lines ratio with 
 the metallicity. This manuscript is organized as follows: in Section~\ref{method} the description of the observational 
 data and the photoionization models are presented; in Section~\ref{res}  we present the resulting calibrations, while the 
 discussion and conclusions of the outcome are given in Sections~\ref{disc} and ~\ref{conc}, respectively.

\section{Methodology}
\label{method}

With the aim of obtaining semi-empirical calibrations between the metallicity and the \ion{N}{v}$\lambda$1240/\ion{He}{ii}$\lambda1640$
 and C43 emission-line intensity ratios,  the same 
methodology proposed by \citet{castro17} was adopted, i.e.\ the calibrations are derived through  diagnostic diagrams
containing observational and model predicted line-intensity ratios. In what follow, the observational data
and the photoionization model descriptions are presented.

\subsection{Observational data}
\label{obs}
 
We compiled from the literature fluxes of the \ion{N}{v}$\lambda$1240, \ion{C}{iv}$\lambda$1549,
\ion{He}{ii$\lambda$}1640, and \ion{C}{iii}]$\lambda$1909  emission-lines produced in 
NLRs of type-2 AGNs. The sample of objects is mainly the one considered
by \citet{matsuoka18} with the addition of Seyfert 2 AGNs located at low redshifts ($z \:  \la \: 0.04$).
In this way our sample comprises objects in the  redshift range  $0 \la z \la 4.0$, divided in
 Seyfert 2 (9 objects), type 2 Quasars (6 objects), high-$z$ Radio Galaxies (61 objects)  and    
Radio-quiet type-2 AGNs (1 object).

In Table~\ref{tab0} the identification, redshift, logarithm of the  
\ion{N}{v}$\lambda$1240/\ion{He}{ii}$\lambda$1640, \ion{C}{iii}]$\lambda$1909/\ion{C}{iv}$\lambda$1549 and
  C43=log[(\ion{C}{iv}$\lambda$1549+\ion{C}{iii}]$\lambda$1909)/\ion{He}{ii}$\lambda$1640] emission-line intensity ratios
   as well as the nebular parameters derived 
  (see below) are listed.  The emission-lines are not reddening corrected due to the small effect of dust extinction on
   metallicity and ionization degree determinations obtained from the considered emission-line ratios  \citep{nagao06a}.
     For some objects were not possible to estimate the error in the line ratios since the uncertainties in the measurements 
     of the emission-line fluxes are not given in the works from which the data were compiled.
The \ion{N}{v}$\lambda$1240 flux is only available for about $30\%$ of the objects in our sample.

\begin{table*}
\caption{Data of AGNs from the literature and derived parameters. Object name, redshift, logarithm of the emission-line
 intensity ratios, metallicity (in units of solar metallicity $Z/Z_{\odot}$) and the logarithm of the number of ionizing 
 photons [$\log Q(\rm H)$] derived from the diagnostic diagrams shown in Fig.~\ref{f1}, are listed for each object.
Diag.~1 refers to estimations obtained from  \ion{N}{v}/\ion{He}{ii} vs.\ \ion{C}{iii}/\ion{C}{iv} diagram (Fig.~\ref{f1}: lower panel) 
and Diag.~2 from C43 vs.\ \ion{C}{iii}/\ion{C}{iv} (Fig.~\ref{f1}: upper panel). In the last two columns  the logarithm of the 
stellar mass (in units of the solar mass) taken from \citet{matsuoka18}, and the references to the works from which  
emission-line intensities were compiled are listed.}
\label{tab0}
\begin{tabular}{@{}l@{}ccccc@{  }c@{ }cc@{  }cc@{  }c@{}}
\hline		 
\noalign{\smallskip}                                
                                &                         &                                          &                               &            &            \multicolumn{2}{c}{Diag.~1}                 &                & \multicolumn{2}{c}{Diag.~2}               &                             &                   \\
\cline{6-7}
\cline{9-10}				
 Object                 &   redshift     &  log(\ion{N}{v}/\ion{He}{ii}) &                  C43         & log(\ion{C}{iii}]/\ion{C}{iv})          & $Z/Z_{\odot}$   &       $\log U$                       &                 &   $Z/Z_{\odot}$                     &    $\log U$             &  $\log(\frac{M}{M_{*}})$    & Refs.\\
\noalign{\smallskip}
\hline 
\hline                                                                            
                                                              \multicolumn{12}{c}{Seyfert 2}                                                                                                                                                                                                                                          \\		
\noalign{\smallskip}
NGC\,1068               &        0.004   &       $0.07\pm0.10$            &              $0.60\pm0.08$  &     $-0.33 \pm 0.09$                     &   ---                  &  ---                          &                  &  $0.81_{-0.29}^{+0.34}$            &    $-1.43_{-0.10}^{+0.16}$      &      ---                &    1 \\
NGC\,4507		&        0.012   &       $-0.03\pm0.11$           &              $0.53\pm0.10$  &     $-0.36   \pm 0.12$                   & $0.82_{-0.29}^{+0.96}$ & $-1.39_{-0.15}^{+0.12}$       &                  &  $1.03_{-0.45}^{+0.60}$            &    $-1.42_{-0.13}^{+0.20}$      &      ---                &    1 \\
NGC\,5506		&        0.006   &        ---                     &              $0.60\pm0.15$  &     $-0.09  \pm 0.15$                    &    ---                 & ---                           &                  &  $0.44_{-0.12}^{+0.61}$            &    $-1.72_{-0.17}^{+0.17}$      &      ---                &    1 \\
NGC\,7674		&        0.029   &        ---                     &              $0.57\pm0.15$  &     $-0.15 \pm   0.19$                   &    ---                 & ---                           &                  &  $0.60_{-0.07}^{+0.81}$            &    $-1.63_{-0.21}^{+0.24}$      &      ---                &    1 \\
Mrk\,3                 	&        0.014   &       $-0.47\pm0.15$           &              $0.52\pm0.05$  &     $-0.36  \pm  0.06$                   & $2.65_{-1.39}^{+0.45}$ & $-1.44_{-0.05}^{+0.04}$       &                  &  $1.08_{-0.23}^{+0.28}$            &    $-1.42_{-0.07}^{+0.11}$      &      ---                &    1 \\		
Mrk\,573                &        0.017   &       $-0.30\pm0.09$           &              $0.47\pm0.08$  &     $-0.51 \pm  0.09$                    &    ---                 &  ---                          &                  &  $1.45_{-0.40}^{+0.85}$            &    $-1.26_{-0.13}^{+0.12}$      &      ---                &    1 \\		
Mrk\,1388	        &        0.021   &        ---                     &              $0.49\pm0.08$  &     $-0.36  \pm  0.08$                   &   ---                  &  ---                          &                  &  $1.16_{-0.37}^{+0.46}$            &    $-1.43_{-0.09}^{+0.11}$      &      ---                &    1 \\	
MCG-3-34-64	        &        0.017   &       $-0.30\pm0.08$           &              $0.32\pm0.10$  &     $-0.30 \pm  0.11$                    & $1.16_{-0.33}^{+1.40}$ & $-1.50_{-0.09}^{+0.10}$       &                  &  $1.56_{-0.64}^{+1.30}$            &    $-1.52_{-0.10}^{+0.14}$      &      ---                &    1 \\	
NGC\,7674               &        0.029   &          ---                   &              $0.64\pm0.14$  &     $ -0.15 \pm 0.14$                    &   ---                 &  ---                           &                  &  $0.42_{-0.11}^{+0.61}$            &    $-1.66_{-0.15}^{+0.15}$      &      ---                &    2 \\
\noalign{\smallskip}
\hline
                                   \multicolumn{12}{c}{ Type 2 Quasar}                                                                                                                                                                                                                                                                 \\			 
CDFS-031                &     1.603      &            ---                 &              $0.41\pm0.04$  &     $-0.36 \pm 0.06$                     &    ---                &  ---                          &                  &  $1.45_{-0.32}^{+0.34}$            &   $-1.44_{-0.04}^{+0.07}$       &      11.43              &     1 \\
CDFS-057$^{(a)}$        &     2.562      &       $0.04\pm0.08$            &              $0.61\pm0.04$  &     $-0.12 \pm 0.03$                     &   ---                 &  ---                          &                  &  $0.52_{-0.17}^{+0.10}$            &   $-1.68_{-0.06}^{+0.05}$       &      10.67              &     1 \\
CDFS-112a               &     2.940      &       $0.21\pm0.04$            &              $0.34\pm0.05$  &     $-0.52  \pm 0.08$                    & $0.94_{-0.23}^{+0.50}$& $-1.17_{-0.06}^{+0.04}$       &                  &  $2.36_{-0.67}^{+0.95}$            &   $-1.29_{-0.07}^{+0.10}$       &      ---                &     1 \\  
CDFS-153                &     1.536      &         ---                    &              $0.80\pm0.08$  &     $-0.26  \pm 0.05$                    &  ---                  &  ---                          &                  &  $0.26_{-0.06}^{+0.28}$            &   $-1.55_{-0.04}^{+0.12}$       &      ---                &     1 \\  
CDFS-531                &     1.544      &         ---                    &              $0.32\pm0.04$  &     $-0.18  \pm 0.05$                    &  ---                  &  ---                          &                  &  $1.18_{-0.16}^{+0.33}$            &   $-1.62_{-0.05}^{+0.05}$       &      11.70              &     1 \\    
CXO\,52                 &     3.288      &       $-0.45\pm0.10$           &              $0.51\pm0.05$  &     $-0.22  \pm 0.04$                    & $1.13_{-0.48}^{+0.58}$& $-1.58_{-0.04}^{+0.04}$       &                  &  $0.82_{-0.20}^{+0.23}$            &   $-1.56_{-0.04}^{+0.04}$       &      ---                &     1 \\             
\hline
                                 \multicolumn{12}{c}{High-z Radio Galaxy}  \\			 
TN\,J0121+1320          &     3.517      &       ---                      &             $0.21\pm0.02$   &     $0.03  \pm 0.01$                     &    ---                &    ---                        &                  &  $1.02_{-0.07}^{+0.05}$           & $-1.83_{-0.01}^{+0.01}$          &      11.02              &     4    \\
TN\,J0205+2242          &     3.507      &       ---                      &             $0.39\pm0.04$   &     $-0.31 \pm 0.05$                     &   ---                 &    ---                        &                  &  $1.26_{-0.16}^{+0.47}$           & $-1.48_{-0.05}^{+0.04}$          &      10.82              &     4    \\  
MRC\,0316-257           &     3.130      &       ---                      &             $0.30\pm0.02$   &     $0.11  \pm 0.02$                     &   ---                 &   ---                         &                  &  $0.74_{-0.05}^{+0.07}$           & $-1.92_{-0.04}^{+0.03}$          &      11.20              &     4    \\     
USS\,0417-181           &     2.773      &       ---                      &             $0.26\pm0.03$   &     $0.19  \pm  0.04$                    &   ---                 &  ---                          &                  &  $0.63_{-0.06}^{+0.19}$           & $-2.04_{-0.04}^{+0.08}$          &      ---                &     4    \\ 
TN\,J0920-0712          &     2.758      &     $-0.30\pm0.01$             &             $0.41\pm0.01$   &     $-0.23  \pm 0.01$                    & $0.88_{-0.04}^{+0.04}$& $-1.54_{-0.01}^{+0.01}$       &                  &  $1.10_{-0.01}^{+0.01}$           & $-1.56_{-0.01}^{+0.01}$          &      ---                &     4    \\ 
WN\,J1123+3141$^{(a)}$  & 3.221          &     $0.60\pm0.01$              &             $0.61\pm0.01$   &     $-0.93  \pm 0.06$                    & $1.86_{-0.34}^{+1.17}$& $-0.81_{-0.01}^{+0.02}$       &                  &  ---                              & ---                              &     $<11.72$            &     4    \\ 
4C\,24.28$^{(a)}$       &     2.913      &     $0.09\pm0.01$              &             $0.32\pm0.01$   &     $-0.18  \pm 0.02$                    &  ---                  &  ---                          &                  &  $1.17_{-0.05}^{+0.05}$           & $-1.62_{-0.02}^{+0.02}$          &     $<11.11$            &     4    \\ 
USS\,1545-234           &     2.751      &     $0.18\pm0.01$              &             $0.34\pm0.01$   &     $-0.34   \pm 0.02$                   &  ---                  &  ---                          &                  &  $1.70_{-0.16}^{+0.10}$           & $-1.48_{-0.02}^{+0.02}$          &      ---                &     4    \\ 
USS\,2202+128           &      2.705     &     $-0.25\pm0.05$             &             $0.53\pm0.01$   &     $-0.38   \pm  0.01$                  & $1.44_{-0.23}^{+0.31}$& $-1.43_{-0.02}^{+0.02}$       &                  &  $1.05_{-0.07}^{+0.08}$           & $-1.40_{-0.02}^{+0.02}$          &     11.62               &     4    \\ 
USS\,0003-19            &      1.541     &       ---                      &                0.37         &        $-0.23$                           &   ---                 & ---                           &                  &   1.16                            & $-1.56$                          &     ---                 &     5    \\ 
MG\,0018+0940           &       1.586    &      ---                       &                0.60         &         0.03                             &  ---                  & ---                           &                  &   0.34                            & $-1.89$                          &     ---                 &     5    \\ 
MG\,0046+1102           &       1.813    &      ---                       &                0.42         &         0.07                             &  ---                  & ---                           &                  &   0.61                            & $-1.89$                          &     ---                 &     5    \\ 
MG\,0122+1923           &       1.595    &      ---                       &                0.22         &         0.00                             &  ---                  & ---                           &                  &   1.05                            & $-1.80$                          &     ---                 &     5    \\ 
USS\,0200+015           &      2.229     &      ---                       &                0.40         &        $-0.02$                           &  ---                  & ---                           &                  &    0.75                           & $-1.78$                          &     ---                 &     5    \\
USS\,0211-122           &      2.336     &        0.12                    &                0.40         &        $-0.40$                           & 0.62                  & $-1.30$                       &                  &   1.58                            & $-1.42$                          &   $<11.16$              &     5    \\
USS\,0214+183           &      2.130     &        ---                     &                0.42         &        $-0.22$                           & ---                   &  ---                          &                  &    1.04                           & $-1.58$                          &     ---                 &     5    \\
MG\,0311+1532           &       1.986    &        ---                     &                0.43         &        $-0.20$                           & ---                   &  ---                          &                  &    0.93                           & $-1.57$                          &     ---                 &     5    \\ 
BRL\,0310-150           &       1.769    &       ---                      &                0.57         &        $-0.30$                           &  ---                  &  ---                          &                  &   0.86                            & $-1.47$                          &     ---                 &     5    \\ 
USS\,0355-037           &        2.153   &        ---                     &                0.13         &        $-0.06$                           &  ---                  &  ---                          &                  &   1.34                            & $-1.73$                          &     ---                 &     5    \\   
USS\,0448+091           &        2.037   &       ---                      &                0.44         &         $0.35$                           &  ---                  &  ---                          &                  &    0.25                           & $-2.25$                          &     ---                 &     5    \\  
USS\,0529-549$^{(b)}$     &      2.575     &       ---                      &                0.56         &         $0.65$                           & ---                   & ---                         &                  &    ---                            &  ---                             &     11.46               &     5    \\  
4C\,+41.17              &        3.792   &       ---                      &                0.60         &        $-0.16$                           & ---                   & ---                           &                  &   0.56                            & $-1.63$                          &     11.39               &     5    \\ 
USS\,0748+134           &        2.419   &       ---                      &                0.34         &        $-0.07$                           &  ---                  &  ---                          &                  &  0.92                             & $-1.71$                          &     ---                 &     5    \\ 
USS\,0828+193           &        2.572   &       ---                      &                0.31         &         0.02                             &  ---                  &  ---                          &                  & 0.85                              & $-1.82$                          &   $<11.60$              &     5    \\ 
BRL\,0851-142           &        2.468   &        ---                     &                0.33         &        $-0.32$                           &  ---                  &  ---                          &                  & 1.71                              & $-1.50$                          &     ---                 &     5    \\ 
TN\,J0941-1628          &        1.644   &      ---                       &                0.76         &        $-0.20$                           &  ---                  &  ---                          &                  & 0.27                              & $-1.63$                          &     ---                 &     5    \\ 
USS\,0943-242           &        2.923   &      $-0.20$                   &                0.36         &        $-0.22$                           &   0.57                &  $-1.55$                      &                  & 1.17                              & $-1.57$                          &     11.22               &     5    \\ 
MG\,1019+0534           &        2.765   &      $-0.56$                   &                0.25         &        $-0.32$                           &   2.83                &  $-1.47$                      &                  & 2.20                              & $-1.49$                          &     11.15               &     5    \\ 
TN\,J1033-1339          &        2.427   &      ---                       &                0.57         &        $-0.51$                           &  ---                  &   ---                         &                  & 1.12                              & $-1.24$                          &     ---                 &     5    \\ 
TN\,J1102-1651          &        2.111   &      ---                       &                0.20         &          0.04                            &  ---                  &    ---                        &                  & 1.01                              & $-1.84$                          &     ---                 &     5    \\ 
USS\,1113-178$^{(b)}$  &        2.239   &      ---                       &                0.80         &          0.21                             &   ---                 &    ---                        &                  & ---                               & ---                              &     ---                 &     5    \\ 
3C\,256.0               &       1.824    &      $-0.59$                   &                0.24         &        $-0.08$                           &  0.81                 &  $-1.71$                      &                  & 1.16                              & $-1.71$                          &     ---                 &     5    \\ 
USS\,1138-262           &       2.156    &     ---                        &                0.20         &          0.21                            &  ---                  &  ---                          &                  & 0.76                              & $-2.03$                          &   $<12.11$              &     5    \\ 
BRL\,1140-114           &       1.935    &     ---                        &                0.50         &        $-0.22$                           &  ---                  &  ---                          &                  & 0.84                              & $-1.56$                          &     ---                 &     5    \\    
4C\,26.38               &       2.608    &      ---                       &                0.29         &        $-0.56$                           &   ---                 & ---                           &                  & 3.06                              & $-1.23$                          &     ---                 &     5    \\
MG\,1251+1104           &        2.322   &      ---                       &                0.43         &         0.23                             &   ---                 & ---                           &                  &  0.36                             & $-2.10$                          &     ---                 &     5    \\
WN\,J1338+3532          &        2.769   &      ---                       &                0.06         &         0.22                             &   ---                 & ---                           &                  & 0.86                              & $-2.08$                          &     ---                 &     5    \\
\hline
\end{tabular}
\end{table*}

\begin{table*}
\setcounter{table}{0}
\caption{-$continued$}
\vspace{0.3cm}
\label{tab0}
\begin{tabular}{@{}l@{}ccccc@{  }c@{ }cc@{  }cc@{  }c@{}}
\hline		 
\noalign{\smallskip}                                
                                &                  &                                          &                                &             &\multicolumn{2}{c}{Diag.1}&  & \multicolumn{2}{c}{Diag.2}            &                             &                   \\
\cline{6-7}
\cline{9-10}				
 Object             &    redshift        &  log(\ion{N}{v}/\ion{He}{ii})  &              C43       &   log(\ion{C}{iii}]/\ion{C}{iv})        & $Z/Z_{\odot}$   & $\log U$ &     & $Z/Z_{\odot}$        &    $\log U$       &  $\log(\frac{M}{M_{*}})$    &   Refs.\  \\
 \noalign{\smallskip} 
\hline
\hline                                                                                                        
													
													\multicolumn{12}{c}{High-z Radio Galaxy}                                                                                                                     \\
\noalign{\smallskip} 
MG\,1401+0921       &      2.093          &        ---                     &           0.17          &             $-0.08$                    &    ---           &   ---  &     &     1.27             &    $-1.72$        &                                  & 5    \\
3C\,294             &      1.786          &     $-0.69$                    &           0.34          &              0.07                      &    ---           & ---    &     &     0.72             &    $-1.89$        &       11.36                      & 5    \\
USS\,1410-001       &      2.363          &    $-0.17$                     &           0.24          &             $-0.47$                    &    1.78          & $-1.34$&     &     3.05             &    $-1.32$        &      $<11.41$                    & 5    \\
USS\,1425-148       &      2.349          &     ---                        &           0.15          &             $-0.36$                    &   ---            &  ---   &     &     3.37             &    $-1.42$        &      ---                         & 5    \\
USS\,1436+157       &      2.538          &     ---                        &           0.64          &             $-0.25$                    &   ---            &   ---  &     &     0.60             &    $-1.51$        &      ---                         & 5    \\
3C\,324.0           &      1.208          &    ---                         &           0.42          &             $-0.02$                    &   ---            &  ---   &     &     0.70             &    $-1.77$        &      ---                         & 5    \\
USS\,1558-003       &      2.527          &    ---                         &           0.36          &             $-0.35$                    &   ---            &  ---   &     &     1.65             &    $-1.47$        &       $<11.70$                   & 5    \\
BRL\,1602-174       &      2.043          &     ---                        &           0.42          &             $-0.56$                    &    ---           & ---    &     &     1.85             &    $-1.25$        &      ---                         & 5    \\
TXS\,J1650+0955     &      2.510          &     ---                        &           0.21          &             $-0.42$                    &    ---           & ---    &     &     3.15             &    $-1.37$        &      ---                         & 5    \\
8C\,1803+661        &      1.610          &    ---                         &           0.44          &             $-0.44$                    &   ---            &  ---   &     &     1.46             &    $-1.35$        &      ---                         & 5    \\
4C\,40.36           &      2.265          &   ---                          &           0.33          &             $-0.02$                    &   ---            &  ---   &     &     0.87             &    $-1.77$        &       11.29                      & 5    \\
BRL\,1859-235       &      1.430          &   ---                          &           0.24          &               0.14                     &   ---            &  ---   &     &     0.81             &    $-1.95$        &      ---                         & 5    \\
4C\,48.48           &      2.343          &   ---                          &           0.38          &             $-0.33$                    &   ---            &  ---   &     &     1.52             &    $-1.49$        &      ---                         & 5    \\
MRC\,2025-218$^{(a)}$ &    2.630          &       0.24                     &           0.67          &               0.14                     &   ---            &  ---   &     &     ---              &     ---           &       $<11.62$                   & 5    \\
TXS\,J2036+0256     &      2.130          &   ---                          &           0.41          &               0.30                     &   ---            &  ---   &     &     0.35             &    $-2.17$        &      ---                         & 5    \\
MRC\,2104-242       &      2.491          &    ---                         &           0.53          &              $-0.15$                   &   ---            & ---    &     &     0.66             &    $-1.62$        &       11.19                      & 5    \\
4C\,23.56           &      2.483          &       $-0.04$                  &           0.34          &              $-0.21$                   &   ---            &---     &     &     1.17             &    $-1.59$        &       11.59                      & 5    \\
MG\,2121+1839       &      1.860          &   ---                          &           0.74          &              $-0.34$                   &   ---            & ---    &     &     0.52             &    $-1.40$        &       ---                        & 5    \\
USS\,2251-089       &      1.986          &   ---                          &           0.56          &              $-0.34$                   &   ---            &  ---   &     &     0.92             &    $-1.43$        &       ---                        & 5    \\
MG\,2308+0336$^{(a)}$ &      2.457        &       0.16                     &           0.44          &              $-0.14$                   &   ---            &  ---   &     &     0.85             &    $-1.64$        &       ---                        & 5    \\
4C\,28.58$^{(b)}$     &      2.891        &  ---                           &           0.11          &               0.77                     &   ---            &  ---   &     &     0.24             &    $-2.71$        &       11.36                      & 5    \\
MP \,J0340-6507     &      2.289          &   ---                          &      $0.18\pm0.11$      &          $0.22 \pm 0.16$               &   ---            &  ---   &     &     $0.77_{-0.36}^{+0.47}$  &  $-2.05_{-0.19}^{+0.19}$&        ---          & 6   \\
TN\,J1941-1951      &      2.667          &   ---                          &       $0.43\pm0.12$     &          $-0.65 \pm  0.22$             &   ---            &  ---   &     &     $2.08_{-1.03}^{+0.05}$  &  $-1.18_{-0.20}^{+0.29}$&        ---          & 6   \\

MP\.J2352-6154      &      1.573          &   ---                          &      $0.39\pm0.06$      &          $-0.36 \pm 0.08$              &  ---             & ---    &     &     $1.51_{-0.43}^{+0.56}$  &  $-1.46_{-0.06}^{+0.11}$&        ---          & 6   \\
\hline
                                                                              \multicolumn{12}{c}{Radio-quiet type-2 AGNs}                                                                                                                                          \\
COSMOS\,05162      &      3.524           &    ---                         &        $0.78\pm0.05$  &         $-0.46  \pm 0.07$                &   ---            &  ---   &      &   $0.46_{-0.17}^{+0.21}$   &  $-1.25_{-0.13}^{+0.16}$&        10.50        & 7   \\   

\hline
\end{tabular}
\begin{minipage}[c]{2\columnwidth}
$^{(a)}$Observational data located out of the region occupied by the model results in the \ion{N}{v}/\ion{He}{ii} vs.\ \ion{C}{iii}/\ion{C}{iv} diagram (see Fig.\ref{f1}, lower panel). $^{(b)}$Observational data located out of the region occupied by the 
model results in the C43 vs.\ \ion{C}{iii}/\ion{C}{iv} diagram (see Fig.\ref{f1}, upper panel).
References: (1) \citet{nagao06a}, (2) \citet{kraemer94}, (3) \citet{diaz88}, (4) \citet{matsuoka09}, 
(5) \citet{debreuck00}, (6) \citet{bornancini07}, (7) \citet{matsuoka18}. 
\end{minipage}
\end{table*}

\begin{figure}
\centering
\includegraphics[angle=-90,width=1\columnwidth]{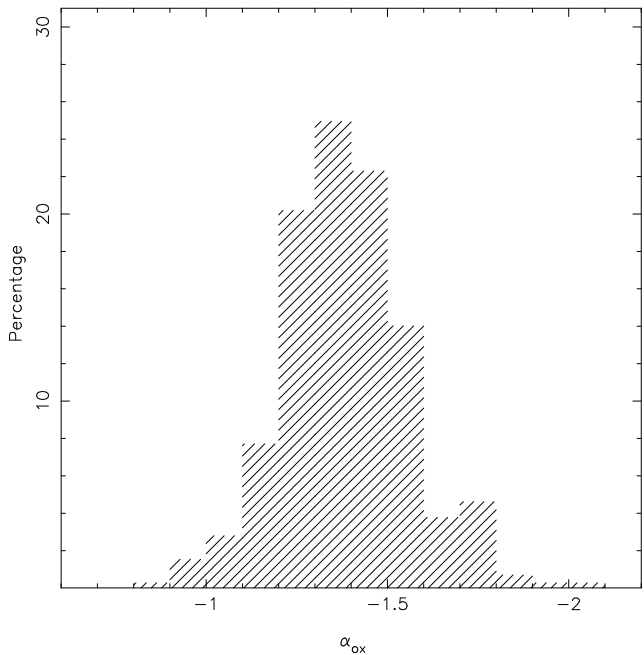}
\caption{Histogram containg the distribution of the spectral index $\alpha_{ox}$   derived by \citet{miller11} for about 700
radio-intermediate and radio-loud Quasars.}
\label{fmil}
\end{figure} 

\begin{figure*}
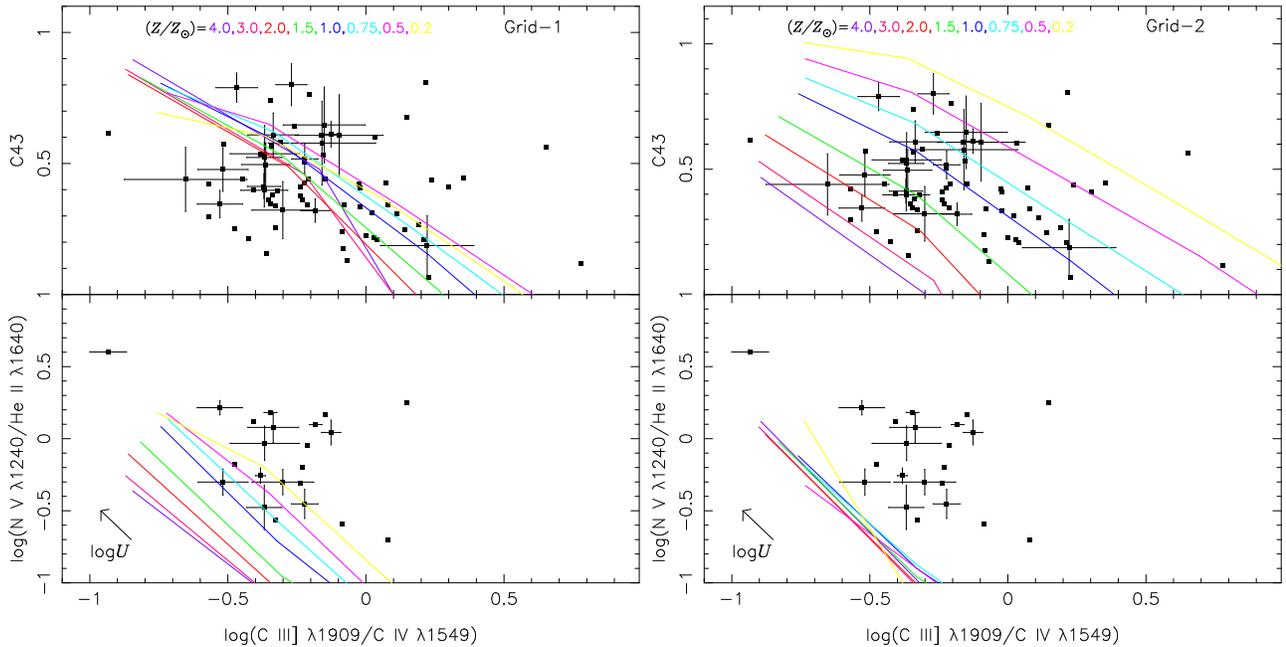

\centering
\includegraphics[angle=-90,width=1\columnwidth]{analis1n.eps}
\includegraphics[angle=-90,width=1\columnwidth]{analis2n.eps}
\caption{Diagrams containing observational data and photoionization model results. Points represent the observational
data listed in Table~\ref{tab0}. Curves represent the photoionization model results described in Sect.~\ref{model},
where  solid lines connect models with equal metallicity ($Z$) as indicated. Left pannel, photoionization
model results assuming the N-O  (Eq.~\ref{eq0a}) and C-O  (Eq.~\ref{eq0b}) relations  proposed by \citet{dors17} and \citet{dopita06}, respectively.
The  arrow indicates the direction in which the number of ionizing photons increases in the models.
Rigth panel: Such as left panel but assuming the relation N-$Z$ relation (Eq.~\ref{eq0bc}) proposed by \citet{hamann93}
and $\rm C/H=(Z/Z)_{\odot} \times (C/H)_{\odot}$.}
\label{f00}
\end{figure*} 
  
\subsection{Photoionization models}
\label{model}
  
The photoionization models were built using the   {\sc Cloudy} code version 17.00 \citep{ferland17} 
in order to compare the predicted UV  emission-line intensity ratios  with those measured for the type-2 AGNs in our sample. 
The range of the input values for the photoionization model parameters are similar to the ones
 used by \citet{dors18} and in what follows we  described them.  
 
 The predicted emission line intensities of our models for the NLRs of type-2 AGNs
are mainly driven by four parameters: 
the ionization parameter ($U$), the metallicity ($Z$), the electron density ($N_{\rm e}$) and the shape of the  Spectral Energy Distribution (SED). 
$U$ is defined as  $U= Q_{{\rm ion}}/4\pi R^{2}_{\rm in} N\, c$, where $ Q_{\rm ion}$  
is the number of hydrogen ionizing photons emitted per second
by the ionizing source, $R_{\rm in}$  is  the distance from the ionization source to the inner surface
of the ionized gas cloud (in cm), $N$ is the  particle density (in $\rm cm^{-3}$), and $c$ is the speed of light.
Excitation differences owing to variations in the mass of the gas phase, or in the geometrical conditions covering a wide range of possible
scenarios, are covered by photoionization models assuming variations of the ionization parameter \citep{perezmontero14}.
We consider $U$  in the range $ -3.5 \: \lid \: \log U \: \lid \: -0.5$, with a step of 0.5 dex; almost the same values considered by \citet{feltre16}
 and \citet{castro17}.

Regarding the electron density, \citet{dors14} found that the NLRs of Seyfert 2 galaxies present gas with low electron density values, 
$N_{\rm e} \: \la \:  3\,000   \: \rm cm^{-3}$
\citep[see also][]{revalski18a, revalski18b}. We assumed in our models a constant electron density
along the radius and equal to $N_{\rm e}= 500 \:\rm cm^{-3}$, 
 an averaged value derived by \citet{dors14}. The outer radius ($R_{\rm out}$) was choosen as the one 
for which the electron temperature reachs 4\,000 K,
since cooler gas practically does not produce UV emission lines. 
The resulting geometry is plane parallel.
It is worth mentioning that models with different combination of $Q_{\rm ion}$, $R_{\rm in}$ and $N_{\rm e}$ 
but that result in the same $U$ are homologous models with
about the same predicted emission-line intensities \citep{bresolin99}.

The Spectral Energy Distribution (SED)   was considered to be composed by two components:
one representing the Big Blue Bump peaking at $\rm 1 \: Ryd$, and the other  a power law with spectral index $\alpha_x=-1$ 
 representing the non-thermal X-ray radiation. As usual, 
the continuum between 2 keV and 2500\AA\ is described by a power law with the spectral index 
\begin{equation}
\alpha_{ox}= \frac{\log [F(2\: {\rm kev})/F(2500 \: {\rm \AA})]}{\log [\nu(2 \: {\rm keV})/\nu(2500 \:{\rm \AA})]} 
\end{equation}
where $F$ is the flux at the given frequence $\nu$ \citep{tananbaum+79}.
 \citet{miller11} combined observational data  of about 700 radio-intermediate and radio-loud Quasars in a wide range of wavelength 
and derived $\alpha_{ox}$  in the range $-2.0 \: \la \: \alpha_{ox} \: \la \: -0.8$
(see also \citealt{zamorani81}). A histogram containg the distribution of the values found by \citet{miller11} is
presented in Figure~\ref{fmil}. We can see that  the high ($\alpha_{ox} \: > \: -1$) and 
low ($\alpha_{ox} \: < \: -1.8$) values are derived for few ($\sim 3 \%$) objects. The averaged $\alpha_{ox}$ is
$-1.37\pm 0.16$. We consider in the models  $\alpha_{ox}=-1.4$, about the same averaged value calculated from estimations by \citet{miller11}.

 Regarding the metallicity, \citet{dors14} found that the C43 index 
is bi-valuated,  yielding one lower branch for $(Z/Z_{\odot}) \: \la \: 0.2$
and an upper branch for $(Z/Z_{\odot}) \: \ga \: 0.2$. This problem is also present
in the  classical $Z$-$R_{23}$ calibration \citep{pagel79} for star forming regions
and this degeneracy can be broken by using the [\ion{N}{ii}]$\lambda$6584/[\ion{O}{ii}]$\lambda$3727 emission-line ratio \citep{kewley08}.
Due to the small number of  measurements of  UV emission lines in most of the AGN spectra, 
the degeneracy in C43 and \ion{N}{v}/\ion{He}{ii} calibrations can only be eliminated based on physical reasons.
Studies of the NLRs of nearby Seyfert 2 AGNs (e.g. \citealt{thaisa98, groves06, dors15, castro17, thomas18,  revalski18a, revalski18b})
and of objects at high redshift (e.g. \citealt{nagao06a, dors14, matsuoka09, matsuoka18}) have showed that metallicities of AGNs are
generally higher than ($Z/Z_{\odot}$)=0.2. Cosmological simulations 
(see Fig.~10 of \citealt{dors14}) and chemical evolution models \citep{molla08}
do not predict low metallicities for the nuclear regions of galaxies (including AGNs).
Besides, \citet{matsuoka18} derived a relation between the stellar mass of galaxies containing type-2 AGN and their NLR  metallicities, which 
were derived using a comparison between the observational intensities of \ion{C}{iv}/\ion{He}{ii} and \ion{C}{iii}]/\ion{He}{ii} line ratios 
and their photoionization model predicted values.
These authors found two metallicity values for galactic stellar masses log($M_\star$/$M_\odot$) between 11 and 11.5, and they pointed out that 
the lowest metallicity solution ($Z/Z_{\odot} \: \la \:0.2$) is unplausible since a significant 
decrease of metallicity with the stellar mass  would be obtained. As these authors remarked, this behaviour has not been reported for any galaxy
 population at any redshift, and has not been reproduced by any model or cosmological simulation.
Therefore, we adopted only metallicities values higher than ($Z/Z_{\odot}$)=0.2 for our
AGN models, i.e.\  the estimated the metallicities using the upper branch of the relation of the metallicity with the C43  index and  with 
\ion{N}{v}/\ion{He}{ii} emission-line ratio.
Assuming a solar oxygen abundance
12+log(O/H)$_\odot$=8.69 \citep{allendeprieto01}, we considered the following values for the metallicity in 
relation to the solar one ($Z/Z_{\odot}$): 0.2, 0.5, 0.75, 1.0, 1.5,
2.0,  and 4.0.

All elements except nitrogen, carbon and helium are taken to be primary nucleosynthesis elements.
For the helium abundance, we used the relation obtained from chemical abundance determinations of \ion{H}{ii} regions  by  \citet{dopita06}:
\begin{equation}
{\rm He/H}=0.0737+0.024 \: \times (Z/Z_{\odot}).
\end{equation}

 Regarding the nitrogen and carbon secondary nucleosynthesis elements, their abundances in relation to
the oxygen abundance (or metallicity) is poorly known for AGNs. In fact, the unique quantitative nitrogen abundance estimation for AGNs 
is the one performed by \citet{dors17} for a small sample of Seyfert 2 galaxies. 
The predicted C43 and \ion{N}{v}/\ion{He}{ii} ratios are 
 dependent on the assumed C-O and N-O relations, respectively. Hence, the derived metallicities are also dependent on the assumed relations.  
 Thus we performed several simulations in order to try to estimate these relations.

Firstly,  we built a grid of models assuming a fixed value of $\alpha_{ox}=-1.4$ and $N_{\rm e}=500  \: \rm cm^{-3}$,
while  $Z$ and $\log U$ ranging in the values described above. In this grid, the N-O relation  obtained by \citet{dors17} 
\begin{equation}
\label{eq0a}
\rm \log(N/H)=1.05 (\pm 0.09) \times [\log(O/H)] -0.35 (\pm 0.33)
\end{equation} 
for nearby Seyfert 2 AGNs, and  the C-O relation 
\begin{equation}
\label{eq0b}
{\rm (C/H)}=6.0 \: \times 10^{-5} \times (Z/Z_{\odot}) + 2.0 \: \times 10^{-4} \times (Z/Z_{\odot})^{2}
\end{equation}
obtained by \citet{dopita06}  for \ion{H}{ii} regions were assumed.  This grid is refered as Grid-1.

In Figure~\ref{f00} (left panels)  the  diagrams  
\ion{N}{v}/\ion{He}{ii} versus \ion{C}{iii}]/\ion{C}{iv} (lower panel) and C43 versus 
 \ion{C}{iii}]/\ion{C}{iv} (upper panel) containing the observational data listed in 
 Table~\ref{tab0} and the photo\-ioni\-za\-tion
 model results (Grid-1) are shown. 
 We  can seen that the predicted intensities of \ion{N}{v}/\ion{He}{ii} and C43 are not in agreement with the observed ones.

\begin{figure}
\centering
\includegraphics[angle=-90,width=1\columnwidth]{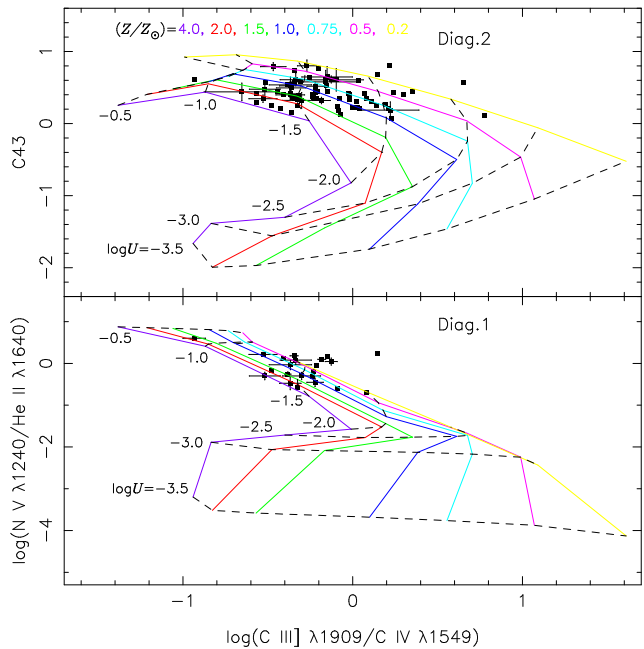}
\caption{Diagrams containing observational data and the best solution for the photoionization model results, i.e.
assuming in the models the relations  $ \mathrm{N/H=\,4.5} \times (\mathrm{Z/Z}_{\odot})^{1.2} \times \mathrm{(N/H)}_{\odot}$
and  $ \mathrm{C/H=} (\mathrm{Z/Z}_{\odot}) \times \mathrm{(C/H)}_{\odot}$ (see Sect.~\ref{model}).
Points represent the observational
data listed in Table~\ref{tab0}. Curves represent the photoionization model results described in Sect.~\ref{model}, where 
solid lines connect models with equal metallicity ($Z$) and dashed lines connect models with equal number of ionizing photons, as indicated.}
\label{f1}
\end{figure}

Secondly,  we adopt a linear scale between the carbon and oxygen abundances
 \begin{equation}
{\rm (C/H)}=(Z/Z_{\odot}) \times {\rm (C/H)_{\odot}}.
\end{equation}
For the nitrogen, the relation
\begin{equation}
\label{eq0bc}
{\rm (N/H)}=\left(Z/Z_{\odot}\right)^{2} \times ({\rm N/H})_{\odot},
\end{equation} 
given by \citet{hamann93} was considered. This relation for the broad-line gas belonging to Quasi Stellar Objects (QSOs) was 
derived applying spectral synthesis and chemical enrichment models \citep[see also][]{hamann02}.
 We assumed $\rm \log(N/H)_{\odot}=-4.07$ and $\rm \log(C/H)_{\odot}=-3.61$
taken from  \citet{holweger01} and \citet{allendeprieto02}, respectively.
The results of this second set of photoionization models (refered as Grid-2) are also shown in 
Fig.~\ref{f00} (rigth panels),
where it can be seen that with the exception of few points,  the carbon lines ratios (upper panel) 
are well reproduced by the models but they fail 
to reproduce the observational \ion{N}{v}/\ion{He}{ii} (lower panel).
Therefore,  we performed a new grid of models assuming 
$$ \mathrm{N/H=\,A} \times (\mathrm{{\it Z/Z}}_{\odot})^{n} \times \mathrm{(N/H)}_{\odot} $$ 
and A  ranging from 1 to 5.0 and  $n$ ranging from  1 to 2. 
The results of these new grids were compared with the  observational line ratios in two diagrams (not shown):
 \ion{N}{v}/\ion{He}{ii}  and C43 versus \ion{C}{iii}]/\ion{C}{iv}. 
 We  found that A=4.5 and n=1.2 produce model results com\-pa\-ti\-ble to 
the  observational data, in the sense that the model results  fit most of the observational data. 
The results of this grid are shown in Fig.~\ref{f1}. It must be noted that the    \ion{N}{v}/\ion{He}{ii} 
preditions are about independent of the carbon abundance assumed
in the mo\-dels, such as  C43 is independent of the nitrogen abundances.

\begin{figure}
\centering
\includegraphics[angle=-90,width=1.0\columnwidth]{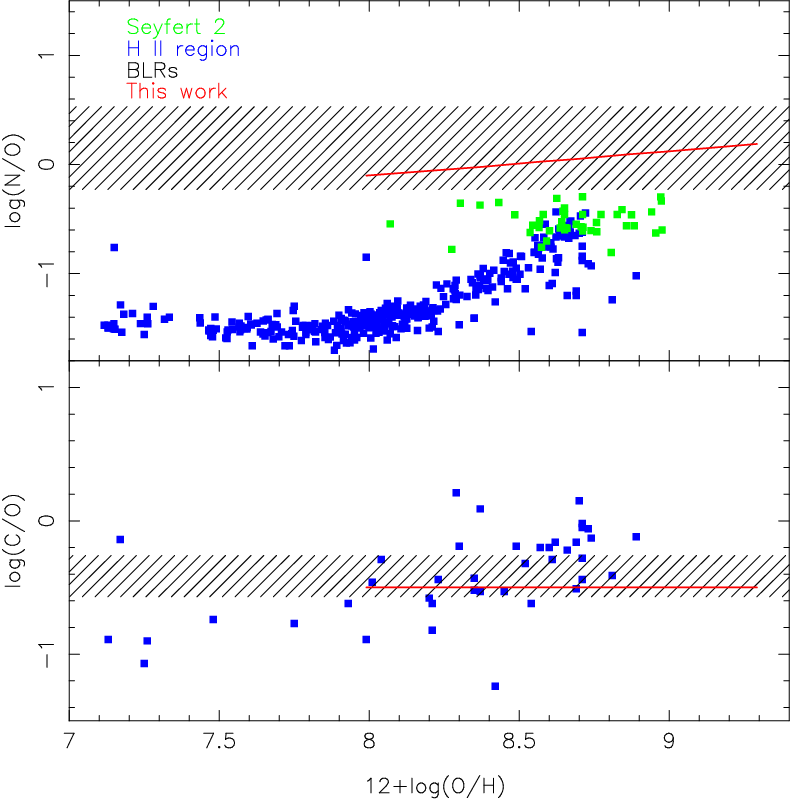}
\caption{ log(N/O) and log(C/O) vs.\ 12+log(O/H) for different kind of
objects, as indicated.
Green points represent abundance ratios for NLRs of Seyfert 2 AGNs at $z<0.1$ from \citet{dors17}. Blue points represent direct estimations of abundance ratios
for star-forming regions taken from \citet{esteban02, esteban09, esteban14}, \citet{jorge07}, \citet{angel07}, \citet{berg16},
and \citet{pilyugin16}.  Hatched areas represent the range of log(N/O) and log(C/O) abundance values
derived  for BLRs of Quasars and based on detailed photoionization models 
by \citet{uomoto84}. Red lines represent the relations assumed in our models.}
\label{f6a}
\end{figure}

 In Fig.~\ref{f6a}, top panel, we compare the logarithm of N/O abundance ratio as a function of 12+log(O/H) (which traces the metallicity) 
assumed in our models ($ \mathrm{N/H=\,4.5} \times (\mathrm{Z/Z}_{\odot})^{1.2} \times \mathrm{(N/H)}_{\odot}$) 
with estimations for NLRs of Seyfert 2 galaxies at $z<0.1$ from \citet{dors17},  
for \ion{H}{ii} regions calculated through the $T_{\rm e}$ method by 
\citet{esteban02, esteban09, esteban14, jorge07}, \citet{angel07}, \citet{berg16},
and \citet{pilyugin16}\footnote{\citet{pilyugin16} used the C-method  \citep{pilyugin12} which produces
similar values than the ones obtained by the $T_{\rm e}$-method.}, 
and  for the Broad Line Regions (BLRs) of Quasars  derived by \citet{uomoto84}  
through detailed photoionization models.
Since  \citet{uomoto84} did not derive the metallicity, i.e.\  only the N/C and O/C abundance ratios were estimated,
his estimations are presented by a hatched area.
We can see that our assumed values are in agreement with those for BLRs of Quasars and they are
higher than the ones for Seyfert 2 and \ion{H}{ii} regions. 

 Also in  Fig.~\ref{f6a}, botton panel, we compare the  
log(C/O) as a function of 12+log(O/H)  assumed in our models, (i.e.\ a fixed 
value $\rm \log(C/O)=-0.50$) with values  for \ion{H}{ii} regions and BLRs of Quasars taken from the same references as in the upper panel.  
Unfortunately, there are no carbon abundance estimation for NLRs of Seyfert 2 galaxies in the literature. We can see that the C-O relation 
assumed in our models is in consonance with the ones for BLRs and \ion{H}{ii} regions. 
A C/O abundance fixed to the solar value was also considered by  \citet{feltre16}, who built a  wide  grid of models in order to 
identify new line-ratio diagnostics to discriminate between gas photoionised by AGNs and 
by hot main sequence stars as well as to estimate the metallicity \citep[see also][]{nagao06a, dors14, matsuoka18}.

\begin{figure*}
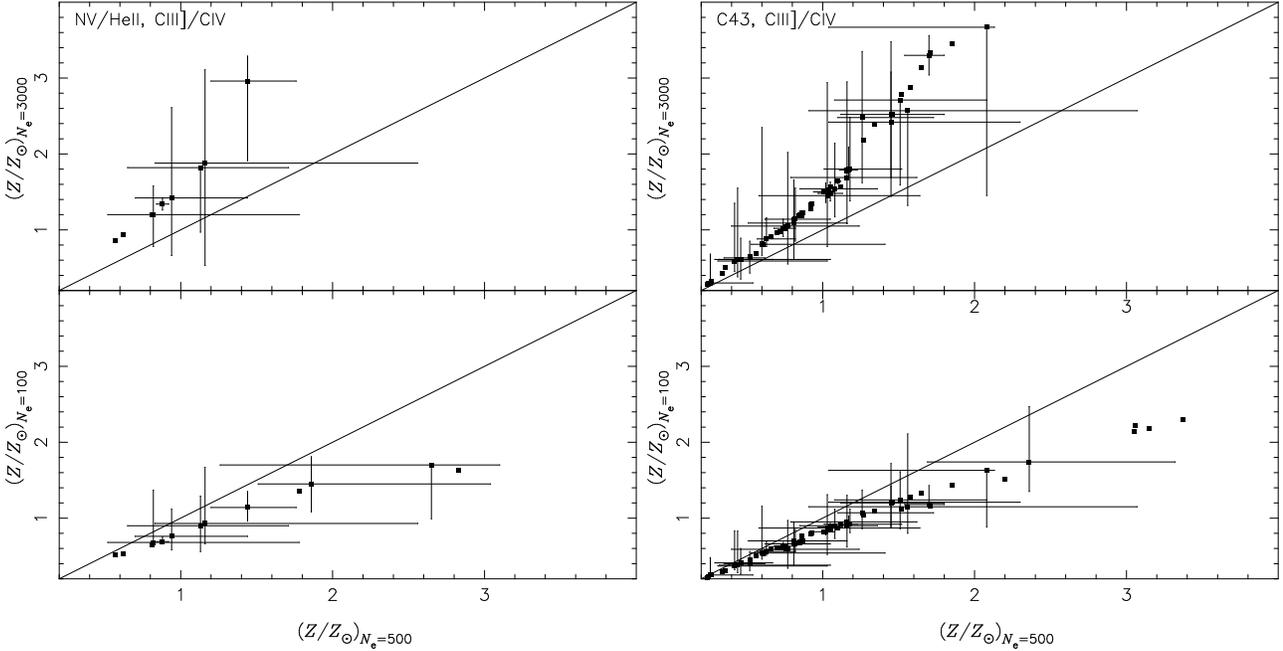

\centering
\includegraphics[angle=-90,width=1\columnwidth]{comp_znv_ne.eps}
\includegraphics[angle=-90,width=1\columnwidth]{comp_zc43_ne.eps}
\caption{Comparison between metallicity ($Z/Z_{\odot}$) estimations for the sample of
objects (see Sect~\ref{obs}) obtained from the diagrams (not shown) \ion{N}{v}/\ion{He}{ii} versus 
\ion{C}{iii}]/\ion{C}{iv} (left panel) and C43 versus \ion{C}{iii}]/\ion{C}{iv} (rigth panel) 
considering grid of models assuming electron density  $N_{\rm e}$= 100, 500 and 3\,000 $\rm cm^{-3}$, as 
indicated in the axis labels.}
\label{var1}
\end{figure*}

As was mentioned above, we consider a fix value for the electron density ($N_{\rm e}=500 \rm \:cm^{3}$). Nevertheless 
the electron density vary in this kind of objects \citep{zhang13, dors14}, producing uncertainties in $Z$  estimations
based on strong emission lines. 
In fact, \cite{feltre16} showed that the location of their models in diagrams containing
 predicted strong optical and ultraviolet emission-line ratios  varies when different $N_{\rm e}$ values
are assumed. 
Also, \citet{nagao06a} found that metallicities based on UV emission lines are larger
when high electron density values ($\sim 10^{5} \: \rm cm^{-3}$) are assumed in 
photoionization models instead of low electron density values ($\la \: 10^{3} \: \rm cm^{-3}$).
 In order to test the $N_{\rm e}$ influence on our $Z$ estimations, firstly, we construct
  grids of models assuming   $N_{\rm e}$= 100, 500 and 3\,000 $\rm cm^{-3}$ 
\citep[the range of values found for NLRs of type 2 AGNs by][]{dors14} constant along
the AGN radius. The results of these grids were compared with the observational
data (see Sect.~\ref{obs}) in  two  diagrams (not shown)
 \ion{N}{v}/\ion{He}{ii} and C43 versus \ion{C}{iii}]/\ion{C}{iv}
 and  the interpolated $Z/Z_{\odot}$ value for each object of the sample was obtained. In Fig.~\ref{var1},
 a comparison of these metallicity estimations is shown. It can be seen that slightly higher metallicity
 values are derived when higher $N_{\rm e}$ values are assumed in the models. However, the discrepancies
 between the estimations are of the order of the uncertainties of these.
Secondly, we consider models with the density ranging along the AGN radius. 
\citet{revalski18a} obtained spatially resolved optical data 
of the  nuclear region of the Seyfert 2 galaxy Markarian 573. 
These authors found a density profile in the NLR of this objects,
with a central electron density peak of about 3\,000 $\rm cm^{-3}$ and a decrease in this value following 
a shallow power law, i.e. $N_{\rm e}\: \propto \: r^{-0.5}$, being $r$ the distance to the AGN center.
We built photoionization models assuming this radial density profile (not shown) and the intensities of UV emission-lines
predicted by these  are about  the same  ones  of  models with $N_{\rm e}$ constant along $r$.
Therefore, electron density variations have an almost negligible influence in the considered line ratios for our estimations.

%
%
%

\begin{figure*}
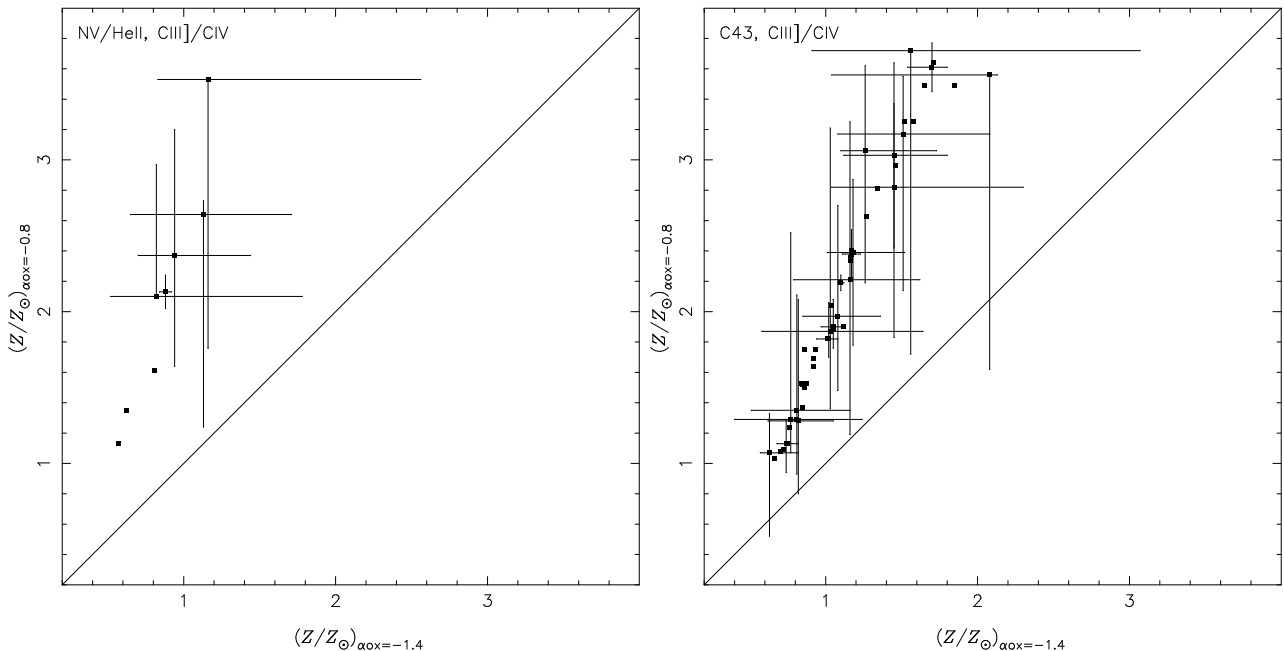

\centering
\includegraphics[angle=-90,width=1\columnwidth]{comp_znv_aox.eps}
\includegraphics[angle=-90,width=1\columnwidth]{comp_zc43_aox.eps}
\caption{Such as Fig.~\ref{var1} but for  grids of models assuming $\alpha_{ox}$ equal to $-1.4$ and $-0.8$.}
\label{var2}
\end{figure*}

Finally, we also explore the dependence of the spectral index $\alpha_{ox}$
 on the  our metallicity estimations.
To do this,   we built  grids of photoionization models assuming three different index values: $\alpha_{ox} = -2.0, -1.4,$ and $-0.8$.
This is the range of $\alpha_{ox}$ values found by the observational study carried out by \citet{miller11} and represented
in Fig.~\ref{fmil}. The SED with $\alpha_{ox}=-2.0$ represents an ionizing source with a very soft spectrum, yielding models
with very low ionization degree and, hence, with emission-line ratios  
largely  discrepant from those of the observational sample. It must be noted that very few ($\la 1\%$) objects studied by \citet{miller11} have $\alpha_{ox}$
 near to $-2.0$. The results of these grids were compared with the observational
data  in  two  diagrams (not shown) \ion{N}{v}/\ion{He}{ii} and C43 versus \ion{C}{iii}]/\ion{C}{iv}
 and  the interpolated $Z/Z_{\odot}$ values were estimated for each object. 
In Fig.~\ref{var2}, a comparison of these metallicity estimations is shown. 
It can be seen that when $\alpha_{ox}=-0.8$ is considered  higher ($\sim 90\%$) metallicity values   are obtained
in comparison with those estimated by  models with $\alpha_{ox}=-1.4$. 
However, as can be seen in Fig.~\ref{fmil},  very hard SEDs ($\alpha_{ox}\: < \: -1.0$)
are derived only for few ($\sim 2\%$) objects and most part of AGNs present $\alpha_{ox}$ around $-1.4$. 
Therefore,  we conclude that our metallicity estimations based on the models with 
$\alpha_{ox}=-1.4$ (see Fig.~\ref{f1}) seem to relate to the real  metallicity of the observed regions.

\citet{thomas18} presented a Bayesian code that implements a general method of comparing observed emission-line
fluxes to photoionization model grids representating AGNs \citep[see also][]{blanc15} and taking into account a large
range of nebular parameters (e.g.\ gas pressure, depletions onto dust grains) and SEDs of the ionizing source. 
Basically, this method determines the probability of that a set of model parameter values is representative of a set of observational 
data of a given object.  Applying this method, \citet{thomas19} found  a correlation between the mass of the host galaxy and the 
metallicity of the AGN \citep[see also][]{matsuoka18} for a sample from the Sloan Digital Sky Survey (SDSS, \citealt{york00}).
\citet{thomas19} pointed out that their sophisticated analysis produces almost an identical mass-metallicity relation to the one 
derived from the $Z$-$N2O2$ calibration obtained by \citet{castro17} and based on similar photoionization models (and methodology) 
applied in the present work (see also \citealt{mignoli19}).
Therefore, metallicity estimations based on symplified photoionization models seem to be reliable.

\section{Results}
\label{res}
 
Diagrams in Fig.~\ref{f1},   \ion{N}{v}/\ion{He}{ii} versus \ion{C}{iii}]/\ion{C}{iv} (Diag.1, lower panel) and C43 versus 
 \ion{C}{iii}]/\ion{C}{iv} (Diag.2, upper panel) were used to calculate the
logarithm of the number of ionizing photons [$\log (Q(\rm H)$]  and the metallicity ($Z/Z_{\odot}$)
for each object in our sample applying a linear interpolations between the
models. These results are listed in Table~\ref{tab0}. The interpolated  parameters together with the observational 
line ratios for each object were used to obtain semi-empirical calibrations. 

\begin{figure*}
\centering
\includegraphics[angle=0.0,width=1.8\columnwidth]{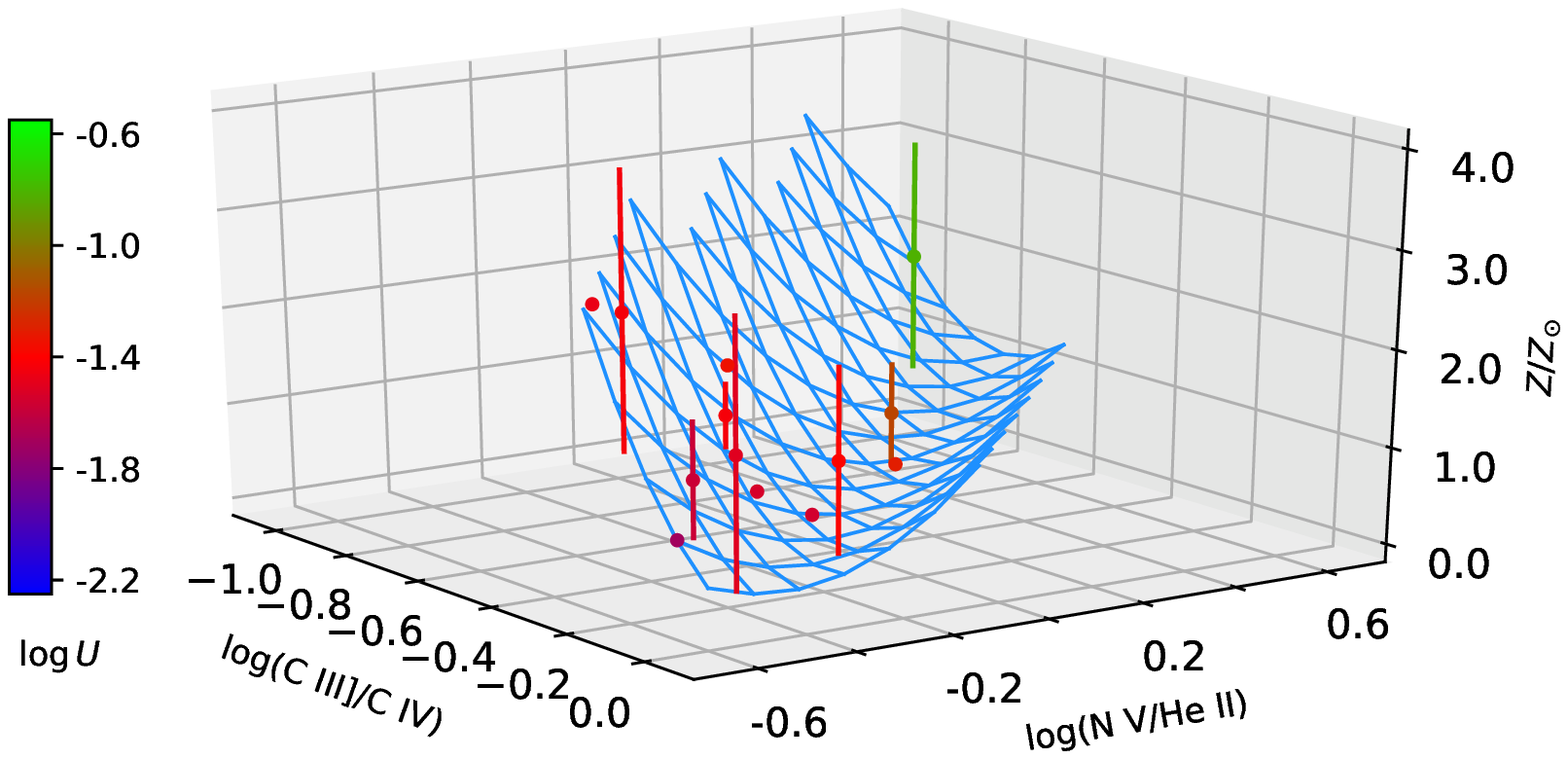}\\
\includegraphics[angle=0.0,width=1.8\columnwidth]{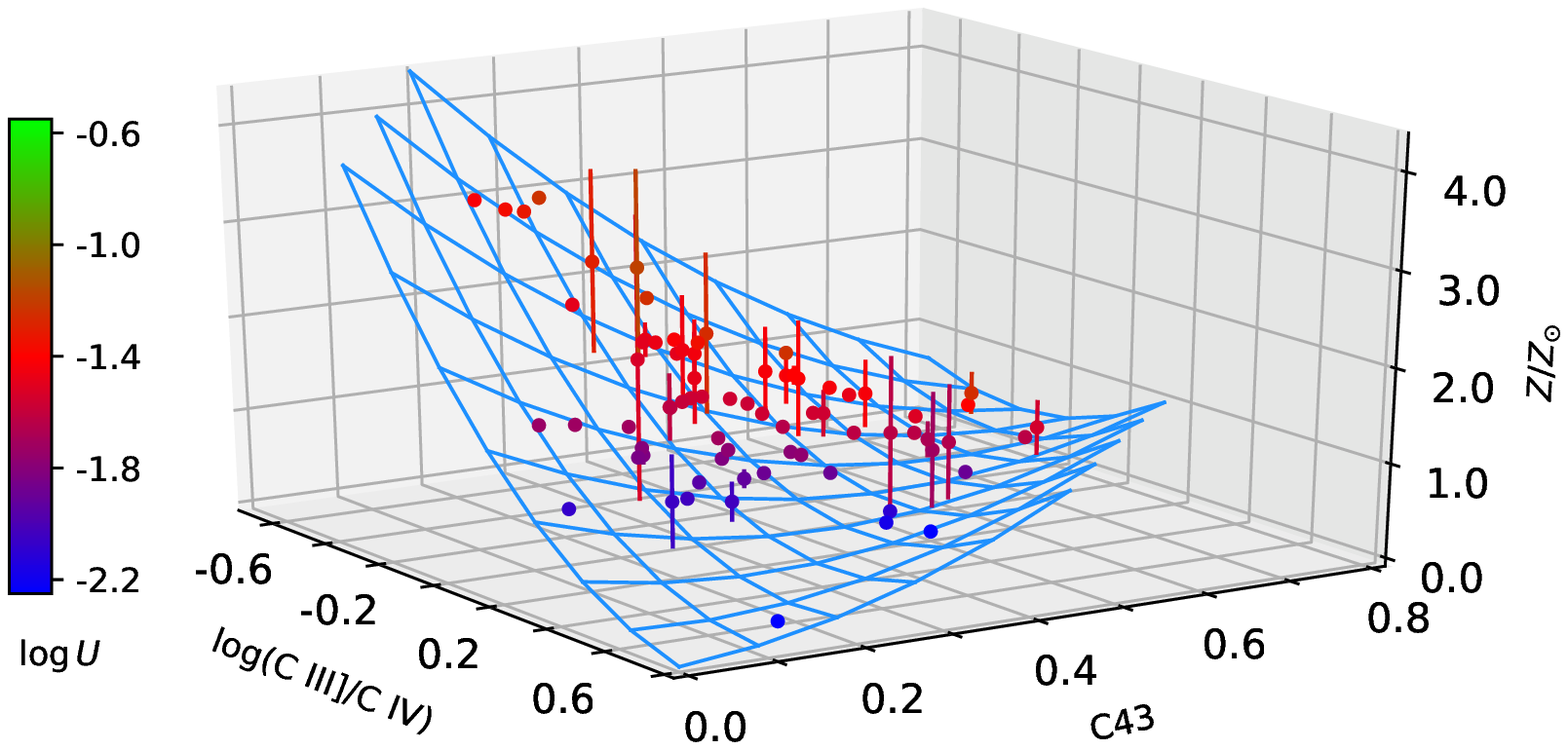}
\caption{Upper panel: bi-dimensional calibrations between  the metallicity ($Z/Z_{\odot}$), log(\ion{N}{v}$\lambda$1240/\ion{He}{ii}$\lambda$1640)
and  log(\ion{C}{iii}]$\lambda$1909/\ion{C}{iv}$\lambda$1549) line ratios.  Points represent metallicity estimations for objects 
listed in Table~\ref{tab0} and  obtained using lower panel of Fig.~\ref{f1} as described in Sect.\,\ref{res}.
The color coding indicates  values of the logarithm of the ionization parameter ($\log U$)
estimated for each object.
The surface represents the best fit to the points and its expresion is given in  Eq.\ref{calibn5}. 
Lower panel: calibration between the metallicity ($Z/Z_{\odot}$), 
C43=log[(\ion{C}{iv}$\lambda$1549+\ion{C}{iii}]$\lambda$1909)/\ion{He}{ii}$\lambda$1640] and 
log(\ion{C}{iii}]$\lambda$1909/\ion{C}{iv}$\lambda$1549) line ratios using metallicities estimations 
obtained using the upper panel of Fig.\,\ref{f1} as described in Sec.\,\ref{res}.
The surface represents the best fit to the points and its expresion is given in Eq.\,\ref{calibc43}.} 
\label{f2}
\end{figure*}

In  Fig.~\ref{f2} (upper panel), the bi-parametric calibration  $(Z/Z_{\odot})=f(\ion{N}{v}/\ion{He}{ii},\ion{C}{iii}]/ \ion{C}{iv})$
is shown. Its expresion is:
\begin{eqnarray}
\begin{array}{lll}

(Z/Z_{\odot})=& (1.48{\scriptstyle \pm0.69})+(12.61{\scriptstyle \pm3.23})x^{2}+(6.28{\scriptstyle \pm0.78})y^{2}  &   \\
              & +(17.66{\scriptstyle \pm3.08}) x\: y +(6.75{\scriptstyle \pm3.03})x                                &    \\
	      & +(5.50{\scriptstyle \pm1.44})y                                                                     &    \\	                                            
\end{array}
 \label{calibn5}
\end{eqnarray}
where $x=\log(\ion{C}{iii}]/\ion{C}{iv})$  and $y=\log(\ion{N}{v}/\ion{He}{ii})$. 
In the same way, the result for the bi-parametric calibration $(Z/Z_{\odot})=f({\rm C}43,\ion{C}{iii}]/ \ion{C}{iv})$ is shown in Fig.~\ref{f2}
 (right panel). The explicit form for this second calibration is:

\begin{eqnarray}
\begin{array}{lll}

(Z/Z_{\odot})=& (2.13{\scriptstyle \pm0.09})+(2.41{\scriptstyle \pm0.19})x^{2}+(4.76{\scriptstyle \pm0.58})C43^{2}   &   \\
              & +(7.79{\scriptstyle \pm0.59}) x \: C43 -(4.64{\scriptstyle \pm0.19})x                                 &    \\
	      & -(5.64{\scriptstyle \pm0.48})C43.                                                                     &	                                            
\end{array}
\label{calibc43}
\end{eqnarray}

\begin{figure}
\centering
\includegraphics[angle=-90,width=1.0\columnwidth]{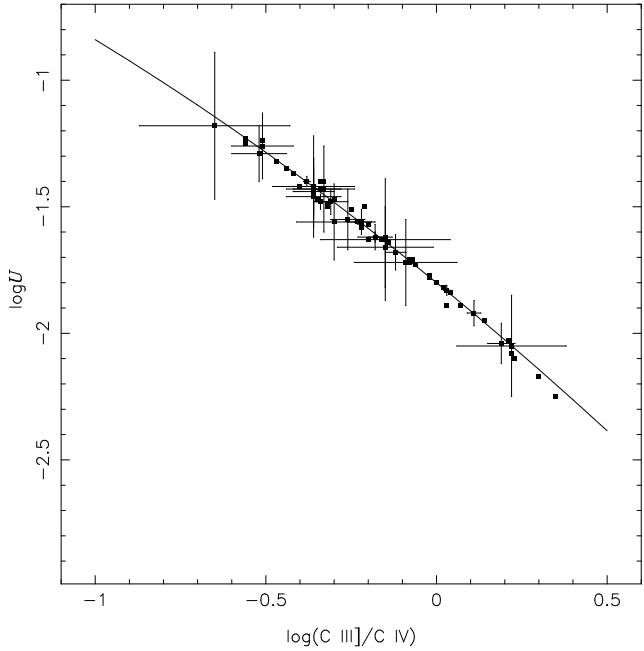}
\caption{ As in Fig.~\ref{f2} but for $\log U$ as a function of the logarithm of 
\ion{C}{iii}]$\lambda$1909/\ion{C}{iv}$\lambda$1549 emission-line ratio. Line represents the best fit.}
\label{f4}
\end{figure}

In Fig.~\ref{f4}, a calibration between the logarithm of the {ionization parameter} 
and the \ion{C}{iii}]/\ion{C}{iv} emission-line ratio is shown. The linear regression obtained is:  
\begin{equation}
\log U= -(0.14 {\scriptstyle\pm 0.02}) \times x^{2} - (1.10 {\scriptstyle\pm 0.01}) \times x - (1.80 {\scriptstyle\pm 0.01}).  
\end{equation}

\begin{figure}
\centering
\includegraphics[angle=-90,width=1.0\columnwidth]{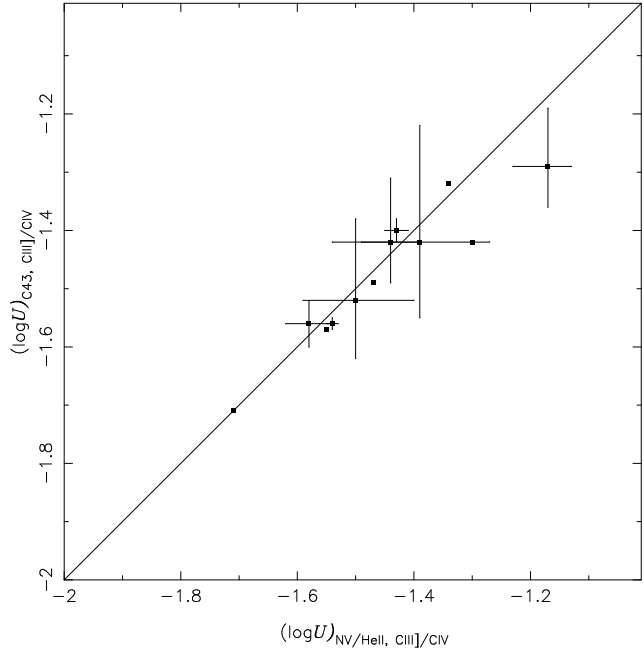}
\caption{ Comparison between $\log U$ obtained with Diag.2  vs.\ $\log U$ obtained with Diag.1  (see 
Fig.~\ref{f1}) and listed in Table~\ref{tab0}. Solid line
represents the one-to-one relation.}
\label{logU}
\end{figure}

In Fig.~\ref{logU} the logarithm of the ionization parameter obtained with the 
diagnostic diagrams presented in Fig.~\ref{f1} (Diag.1 and Diag.2) and 
listed in Table~\ref{tab0} are compared.  This figure shows a very good 
agreement between both estimations.

\section{Discussion}
\label{disc}


\subsection{Metallicity in narrow and board line regions}

The first metallicity estimations in BLRs of AGNs based on ultraviolet emission lines were obtained by 
\citet{shields76} and \citet{davidson77}, who  used the initial  photoionization models (see \citealt{davidson72, shields74, baldwin78})
to investigate which combinations of  line ratios are the most appropriate to estimate
abundance ratios (e.g. N/C and O/C). Later,  \citet{uomoto84} obtained UV spectra of a sample of six Quasars
($1.6 \: < \, z \: < 2.1$) and, fitting photoionization models to the observational line measurements, Uomoto found
 about solar oxygen abundances and a sligth  enhancement of nitrogen abundances for two of these Quasars. Other studies, as for example
\citet{hamann97}, concluded that BLRs, in fact, present super-solar abundances ($Z \: > \: 5 \: Z_{\odot}$) and an enhancement of N abundances
(see also \citealt{peimbert68, alloin73, thaisa90, thaisa91, dietrich03, bradley04}).
More recently, \citet{sameshima17} derived the metallicity, traced by the Fe/H and Mg/Fe  abundance ratios, in  the   BLRs of a sample of about 17\,500 Quasars
($0.7 \: \la \: z \: \la \:1.7$), whose observational data were taken from  the Sloan Digital Sky Survey (SDSS, \citealt{york00}).
To obtain the metallicity, these authors compared observational fluxes and equivalent widths 
of \ion{Fe}{ii}   (calculated by integrating the fitted \ion{Fe}{ii} template in the  $\rm 2200\AA-3090 \AA$ wavelength range)
and \ion{Mg}{ii}$\lambda2798$  lines with photoionization model results. They  found  
that  the majority of the objects have oversolar  metallicities with a median value of $Z \approx 3 Z_{\odot}$, 
reaching up to $(Z/Z_{\odot})\sim 100$ \citep[see also][]{dietrich00, hamann02, batra14}.

\begin{figure}
\centering
\includegraphics[angle=-90,width=1.0\columnwidth]{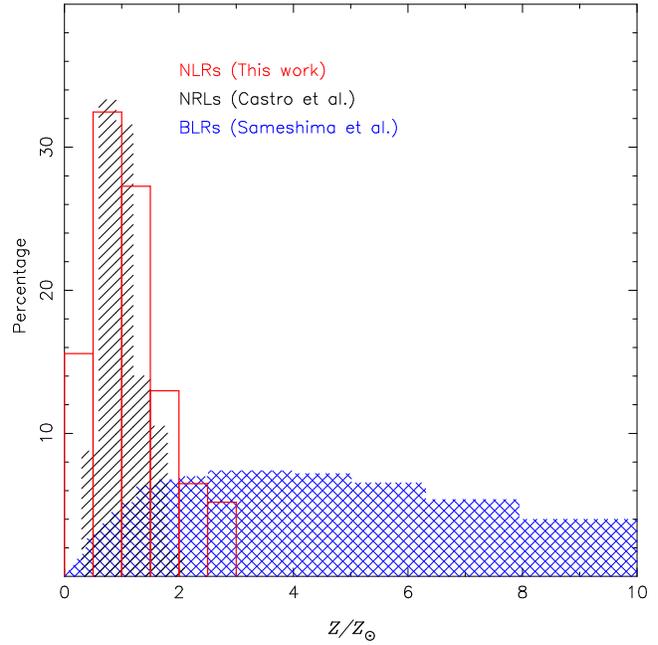}
\caption{Histograms containing NLRs and BLRs metallicity distributions.
Red (empty-filled) histogram represents NLRs metallicity distribution for our sample ($0.04 \: < \: z \: < 4.0$) 
and using the Eq.~\ref{calibc43}. Black histogram (with slant-fill pattern) represents NLRs metallicity distribution
for a sample of Seyfert 2 AGNs ($z \: < 0.1$)
 derived using a semi-empirical calibration proposed by \citet{castro17}.
Cyan histogram (with x-fill pattern) represents  BLRs metallicity distribution for  
 a sample of  Quasars  ($1.6 \: < \, z \: < 2.1$) derived by  \citet{sameshima17}.}
\label{f6}
\end{figure}

In order to  compare our  NLRs $Z$ estimations with those for BLRs, a histogram
containing the metallicity distribution for our sample, derived using the
Eq.~\ref{calibc43},  and the results of \citet{sameshima17} is presented in Fig.~\ref{f6}.
We limited the their estimations to $(Z/Z_{\odot}) \:\lid \: 10$ since few objects
present higher metallicities.
In Fig.~\ref{f6}, the NLR metallicity results derived by  \citet{castro17},  from a semi-empirical calibration between $Z$ and the optical 
ratio $N2O2$=log([\ion{N}{ii}]$\lambda$6584/[\ion{O}{ii}]$\lambda$3727), are also shown.
We found that the  majority   ($\sim 60 \%$) of the NLRs metallicity estimations based on the C43
 are in the  $0.5 \: \la \: (Z/Z_{\odot})  \: \la \: 1.5$ range, with a median value of $<Z/Z_{\odot}> \sim 1.1$.
Very different metallicity distributions are found for NLRs and BLRs, being the later more extended and with a median value higher, 
by a factor of 2-3, than the former.
This difference could be probably due to the fact that broad lines originate in a small region with radius lower
than 1 pc \citep{kaspi00, suganuma} which may evolved more rapidly than the NLRs \citep{matsuoka18}.
As can be seen in Fig.~\ref{f6}, our metallicity estimations follow the same distribution than  those derived by \citet{castro17} for 
the NLRs of Seyfert 2 AGNs, who found 
metallicities in the $0.7 \: \la \: (Z/Z_{\odot})  \: \la \: 1.2$ range for most of their objects.

\subsection{Metallicity estimations from distinct calibrations}

Previous studies of metallicities in AGNs (e.g. \citealt{shemmer02, dietrich03, dhanda07}) and in star-forming regions
(e.g.\ \citealt{kewley08, angel09, dors11}) have shown that different methods or different calibrations of the same index  provide
different metallicity values, with discrepancies of up to 1.0 dex. 

\begin{figure}
\centering
\includegraphics[angle=-90,width=1.0\columnwidth]{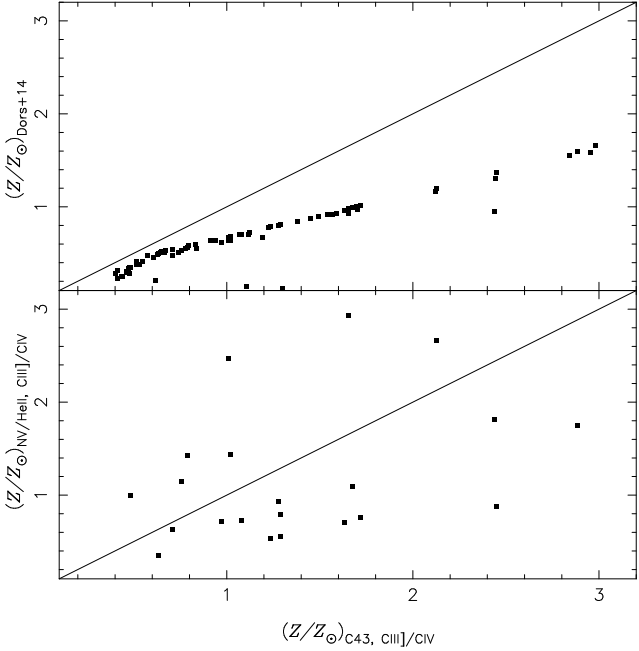}
\caption{Comparison between NLRs metallicity estimations for the objects in our sample 
obtained through distinct calibrations. Solid lines represent the one-to-one relation between the estimations.  
Bottom panel: estimations via 
$Z/Z_{\odot}=f(\ion{N}{v}/\ion{He}{ii},\ion{C}{iii}]/ \ion{C}{iv})$ (Eq.~\ref{calibn5})
vs.\ those via 
$Z/Z_{\odot}=f({\rm C43},\ion{C}{iii}]/ \ion{C}{iv})$ (Eq.~\ref{calibc43}).
Upper panel: estimations via the theoretical calibration $(Z/Z_{\odot})=f({\rm C43},\ion{C}{iii}]/ \ion{C}{iv})$ 
proposed by \citet{dors14} vs.\ those via 
$Z/Z_{\odot}=f({\rm C43},\ion{C}{iii}]/ \ion{C}{iv})$ (Eq.~\ref{calibc43}).
}
\label{f7}
\end{figure}

In order to compare the estimations from our two calibrations, in the bottom panel of Fig.~\ref{f7}, 
$Z$ estimations for the objects in our sample obtained  
from the bi-parametric calibration  $(Z/Z_{\odot})=f(\ion{N}{v}/\ion{He}{ii},\ion{C}{iii}]/ \ion{C}{iv})$
(Eq.~\ref{calibn5})  are plotted against the estimations via $(Z/Z_{\odot})=f(\rm C43,\ion{C}{iii}]/ \ion{C}{iv})$ (Eq.~\ref{calibc43}). 
It can be seen that,  despite the large scattering and the few number of points, the estimations
are in consonance one to each other.

In the top panel of Fig.~\ref{f7}, the estimations  from the semi-empirical calibration $(Z/Z_{\odot})=f({\rm C43},\ion{C}{iii}]/ \ion{C}{iv})$ (Eq.~\ref{calibc43})
are compared to those derived via the theoretical relation  by \citet{dors14}, where it can be seen that
the later one produces $Z$ values higher than the former, mainly for the high metallicity regime. 
This difference is probably due to \citet{dors14} did not deeply explore the dependence between
the metallicity and the $U$-\ion{C}{iii}/\ion{C}{iv} relation. Since the semi-empirical
and bi-parametric calibrations (Eqs.~\ref{calibn5} and \ref{calibc43}) are derived taken into account observational
data, they are, in principle, more robust and confident than the purely theoretical calibration derived by \citet{dors14}.

Finally, we compared our UV estimations for individual objects with the ones based on optical emission lines.
Unfortunately, only for three objects in our sample 
it was possible this comparison: the Seyfert 2 galaxies: NGC\,5506, Mrk\,3, and Mrk\,573. 
We used the optical emission line intensities  of these galaxies, compiled by \citet{dors15}, and 
estimated the  metallicity  through the theoretical 
calibration between $Z$ and $N2O2$ 
index proposed by \citet{castro17}:
\begin{eqnarray}
     \begin{array}{lll}
(Z/Z_{\odot}) & = &\!\!\! 1.08(\pm0.19) \times N2O2^2  +  \\  &+&\!\!\! 1.78(\pm0.07) \times N2O2 +  1.24(\pm0.01)  \\  
     \end{array}
\label{eqcastro}
\end{eqnarray}
and through the first calibration proposed by   \citet[][hereafter SB98]{thaisa98}:
\begin{eqnarray}
       \begin{array}{lll}
{\rm (O/H)}_{{\rm SB98}} \!\!\!\!& = &\!\!\!\!  8.34  + (0.212 \, x) - (0.012 \,  x^{2}) - (0.002 \,  y)  \\  
         \!\!\!\!&+&\!\!\!\! (0.007 \, xy) - (0.002  \, x^{2}y) +(6.52 \times 10^{-4} \, y^{2}) \\  
         \!\!\!\!&+&\!\!\!\! (2.27 \times 10^{-4} \, xy^{2}) + (8.87 \times 10^{-5} \, x^{2}y^{2}),   \\
     \end{array}
\label{sb1}
\end{eqnarray}
where $x$ = [N\,{\sc ii}]$\lambda$$\lambda$6548,6584/H$\alpha$, $y$ = [O\,{\sc iii}]$\lambda$$\lambda$4959,5007/H$\beta$
and 
\begin{equation}
\label{sb2}
(Z/Z_{\odot})_{\rm SB98}=10^{[\rm (O/H)_{SB98} -8.69]}.
\end{equation}

The  estimations using these optical calibrations  are compared with the ones 
via Eqs.~\ref{calibn5} and \ref{calibc43} in Table~\ref{tab1}. We can see that  the
f(C43,\ion{C}{iii}]/\ion{C}{iv}) relation (Eq.~\ref{calibc43}) produces the closest  values  
to the ones from the  optical calibrations (Eqs.\ \ref{eqcastro} and \ref{sb1}).
Nevertheless, it must be noted that only three objects do not constitutes a statistical significant comparison sample.
 
\begin{table}
\centering
\caption{Object names and their corresponding metallicities ($Z/Z_{\odot}$) obtained through the indicated equations.}
\vspace{0.3cm}
\label{tab1}
\scriptsize
\begin{tabular}{@{}l@{ }c@{ }c@{ }c@{ }c@{}}
\hline
\noalign{\smallskip}
          &    \multicolumn{4}{c}{$Z/Z_{\odot}$}         \\
\noalign{\smallskip}
\cline{2-5}	   
\noalign{\smallskip}
  Object                &   f(\ion{N}{V}/\ion{He}{ii},\ion{C}{iii}]/\ion{C}{iv})       & f(C43,\ion{C}{iii}]/\ion{C}{iv})    &     $N2O2$ & SB98	 \\
          &   (Eq.\ \ref{calibn5})                  &  (Eq.\ \ref{calibc43}) &  (Eq.\ \ref{eqcastro}) &  (Eq.\ \ref{sb2}) \\
  NGC\,5506            &     ---                 &      0.47          &       1.15             & 0.81	  \\
  Mrk\,3               &         2.47            &      1.00          &       1.16             & 1.14	  \\
  Mrk\,573             &         1.09  		 &      1.67	      &       1.12             & 0.91  \\		    
\noalign{\smallskip}
\hline
\end{tabular}

\end{table}

\subsection{Mass-Metallicity relation}

\begin{figure}
\centering
\includegraphics[angle=-90,width=1.0\columnwidth]{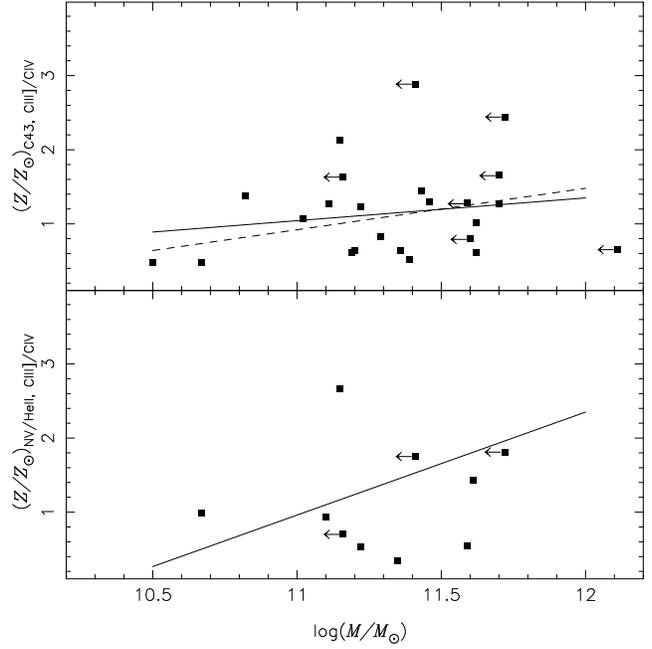}
\caption{$M-Z$ relation for type-2 AGNs. Metallicity values were calculated via Eq.~\ref{calibn5} (bottom panel)
and Eq.~\ref{calibc43} (upper panel).  Solid lines represent the linear regresions (Eqs.~\ref{rel1} and \ref{rel2}).
In the upper panel  the  dashed line represents the linear regression without taking into account the more massive object (Eq.~\ref{rel3}).
Arrows indicate that only the upper limit  of the stellar mass was quoted in the literature (see Table~\ref{tab0}).}
\label{f9}
\end{figure}

Recently, for the first time, \citet{matsuoka18} showed that there is a dependence between the NLR metallicities of type-2 AGNs ($1.2 \: < \: z \: < \:4.0$) 
with the stellar masses of their host galaxies, in the sense that AGNs with higher metallicities are located in more massive galaxies. 
These authors divided the sample in bins of stellar  masses ($M_{*}/M_{\odot}$) and, for each bin, calculated NLRs metallicities by
comparing the \ion{C}{iv}/\ion{He}{ii} and \ion{C}{iii}/\ion{C}{iv} ratios of the averaged observed emission-line fluxes 
with those predicted by photoionization models. \citet{matsuoka18} did not derive an expression for the stellar mass-metallicity 
($M-Z$) relation. In order to verify if our $Z$ estimations indicate the existence of a $M-Z$ relation and to derive an expression for it,  
 the $Z$ estimations obtained from Eqs.~\ref{calibn5} and \ref{calibc43} are plotted in Fig~\ref{f9} (lower and upper panel, respectively) as a 
 function of the stellar mass for those galaxies in our sample with this parameter available in the literature (listed in Table~\ref{tab0}). 
 It must be noted that all these galaxies are in $1.6 \: < \: z \: < \:3.8$.
Although there are few objects with determinations of the stellar masses of the host galaxy, we can see  that the metallicity values estimated 
from Eq.~\ref{calibn5} 
increase with the galaxy mass, obtaining:

\begin{eqnarray}
\label{rel1}
\begin{array}{lcl}
(Z/Z_{\odot})_{\scriptsize{\ion{N}{v}/\ion{He}{ii},\ion{C}{iii}]/\ion{C}{iv}}}\!\!\!\!&=&\!\!\!\![1.39\pm 1.36 \: \times \:  \log(M_{*}/M_{\odot})] \\ 
                                                                                        \!\!\!\!&-&\!\!\!\! (14.33\pm 15.43).
\end{array}
\end{eqnarray} 

The $M-Z$ relation via Eq.~\ref{calibc43} is represented by  

\begin{eqnarray}
\begin{array}{lcl}
(Z/Z_{\odot})_{\scriptsize{{\rm C43},\ion{C}{iii}]/\ion{C}{iv}}}\!\!\!\!&=&\!\!\!\![0.32\pm0.38 \: \times \:  \log(M_{*}/M_{\odot})] \\
\!\!\!\!&-&\!\!\!\! (2.49\pm 4.31).
\end{array}
\label{rel2} 
\end{eqnarray}

The  Pearson Correlation Coefficient (P) for the $M-Z$ relations based on Eqs.~\ref{calibn5} and ~\ref{calibc43}
are 0.32 and 0.18, respectively, indicating a positive but weak correlations.
It can be seen in the upper panel of Fig.\ \ref{f9} that there is only one object with $\log(M_{*}/M_{\odot}) \: > \: 12$ and this mass value
is an upper limit. Removing this point from the fitting, we obtained
\begin{eqnarray}
\label{rel3}
\begin{array}{lcl}
(Z/Z_{\odot})_{\scriptsize{{\rm C43},\ion{C}{iii}]/\ion{C}{iv}}}\!\!\!\!&=&\!\!\!\![0.56\pm0.42 \: \times \:  \log(M_{*}/M_{\odot})] \\
\!\!\!\!&-&\!\!\!\!(-5.56\pm 4.76),
\end{array}
\end{eqnarray}
and P=0.28.

Unfortunately, as can be seen in the lower panel of the Fig.~\ref{f9}, we have a small number of objects 
with metallicities derived from Eq.~\ref{calibn5} and with estimations of their stellar masses. Hence we consider 
that the $M-Z$ relation given by Eq.~\ref{rel1} is not statistically significant, hereafter we will only discuss the $M-Z$ relation given by Eq.~\ref{rel3}.

\begin{figure}
\centering
\includegraphics[angle=-90,width=1.0\columnwidth]{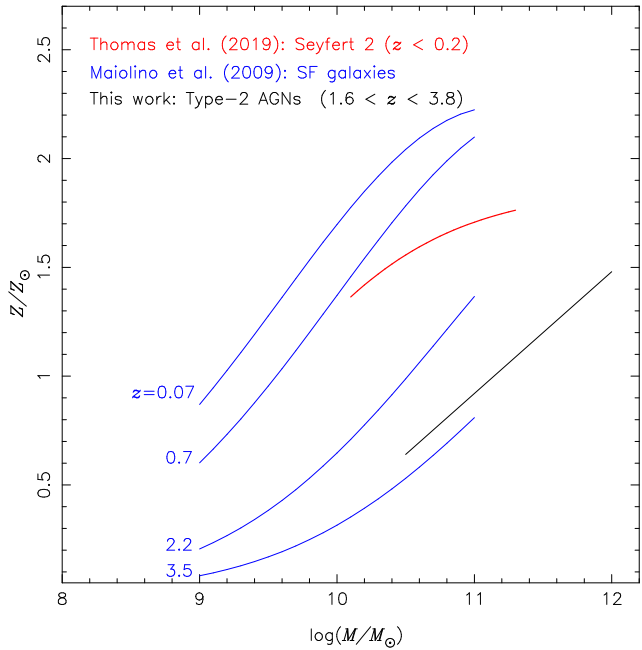}
\caption{ Comparison between our $M-Z$ relation given by Eq.\ref{rel3} and the relations given by \citet{thomas19} 
for local Seyfert 2 AGNs, and by \citet{maiolino08} for SF galaxies at different redshifts ranges (as labeled).}
\label{f10d}
\end{figure}

In Fig.~\ref{f10d}, the $M-Z$ relation given by Eq.~\ref{rel3} is compared
with the relation obtained by \citet{thomas19} for local ($z \: < \: 0.2)$ Seyfert 2 AGNs (in the  
mass interval $10.1 \: < \: [\log(M_{*}/M_{\odot})] \: < \: 11.3$), 
and with $M-Z$ relations for  star-forming galaxies at different 
redshift ranges derived by \citet{maiolino08}.
We can see a difference between our $M-Z$ relation and the one
derived by \citet{thomas19}. Taking into account the very different redshift ranges between the Thomas and collaborators and our samples, 
this difference in the $M-Z$ relation could be explained by the
evolution of this relation with the redshift. 
This interpretation is based on the observed gradual declination of the metallicity with the 
redshift for a given mass in SF galaxies from the Maiolino and collaborators work as can be seen in Fig.~\ref{f10d} \citep[see also][]{savaglio05, erb06, yabe14, ly16}.
It is noticeable that our relation (Eq.~\ref{rel3}) seems to complement the one derived by \citet{maiolino08} for $z\sim 3.5$ towards higher masses.
It is worth emphasizing that a larger sample of AGNs with metallicity estimations 
and with estimations of the stellar mass of the host galaxies, mainly
covering a wider range in masses,  is  needed to improve the $M-Z$ relation.

\section{Conclusion}
\label{conc}

We compared the observational intensities of UV emission lines with results of photoionization
models in order to obtain  two semi-empirical calibrations between the metallicity of the Narrow Line Region
of type-2 Active Galactic Nuclei and the \ion{N}{v}$\lambda$1240/\ion{He}{ii}$\lambda1640$, 
C43=log[(\ion{C}{iv}$\lambda1549$+\ion{C}{iii}]$\lambda1909$)/\ion{He}{ii}$\lambda1640$]
and \ion{C}{iii}]$\lambda1909$/\ion{C}{iv}$\lambda1549$  emission-line ratios. 
 We also derived a metallicity-free  calibration between the 
ionization parameter and the \ion{C}{iii}]$\lambda1909$/\ion{C}{iv}$\lambda1549$ 
emission-line ratio.
We showed that the metallicity in NLRs of AGNs is lower by a factor of 2-3 than the metallicity in BLRs. 
Using the derived calibrations,  we confirmed the recent  result 
of the existence of a relation between the  stellar mass of the host galaxy and its NLR metallicity.
 We were able to derive two 
$M-Z$ relations even though we consider that one of them is not statistically significant due to the 
small number of objects available to derive it.
The adopted $M-Z$ relation was obtained using Type 2 AGNs at $z$ between 1.6 and 3.8. Comparing our 
relation with the ones derived by \citet{maiolino08} for Star Forming galaxies at high redshifts, we 
found that the $M-Z$ relation of Type 2 AGNs seems to complement the sequences towards higher masses following the same trend.

\section*{Acknowledgments}
We are grateful to the anonymous referee for her/his very useful
comments and suggestions that helped us to clarify and improve
this work. OLD and ACK are grateful to FAPESP and CNPq.
AFM is grateful to CAPES. MVC and GFH are grateful to CONICET.
OLD is grateful to Dr. Sameshima and Dr. Miller for making 
their BLR metallicity estimations and  $\alpha_{ox}$ values available, respectively.


\label{lastpage}

\end{document}